\documentclass[lettersize,journal]{IEEEtran}
\usepackage{amsmath,amsfonts}
\usepackage{algpseudocode}
\usepackage{algorithm}
\usepackage{array}
\usepackage[caption=false,font=normalsize,labelfont=sf,textfont=sf]{subfig}
\usepackage{textcomp}
\usepackage{stfloats}
\usepackage{url}
\usepackage{verbatim}
\usepackage{graphicx}
\usepackage{cite}
\usepackage{multirow}
\usepackage{xcolor}
\usepackage{amssymb}

\begin{document}
\bstctlcite{IEEEexample:BSTcontrol}

\title{JOMP: Jointly-Optimized Mixed‑Precision Quantization Across Neural Video Coding Frameworks and Buffering Strategies}

% Assessing Joint-Optimized Mixed-Precision Quantization-Aware Training on Neural Video Codecs

\author{\textcolor{black}{Yu-Hsiang Lin, Ruhan Conceição,~\IEEEmembership{Student Member,~IEEE}, Chun-Hung Wu, \\ Huu-Tai Phung, Tzu-Hsiang Chou, Marcelo Porto,~\IEEEmembership{Senior Member,~IEEE}, \\ Luciano Volcan Agostini,~\IEEEmembership{Senior Member,~IEEE}, Wen-Hsiao Peng,~\IEEEmembership{Fellow,~IEEE}}
\thanks{\textcolor{black}{Yu-Hsiang Lin, Chun-Hung Wu, Huu-Tai Phung, Tzu-Hsiang Chou, and Wen-Hsiao Peng are affiliated with the Department of Computer Science, National Yang Ming Chiao Tung University, Hsinchu, Taiwan (e-mail: wpeng@cs.nctu.edu.tw).}}
\thanks{\textcolor{black}{Ruhan Conceição, Marcelo Porto, and Luciano Volcan Agostini are affiliated with the Graduate Program in Computing, Federal University of Pelotas, Brazil.}}}


\maketitle
%VARIABLE NAMES

%Component of NVC Originally c
\newcommand{\comp}{k}

%Number of components in a NVC
\newcommand{\comptotal}{k}

%Number of components in a NVC
\newcommand{\layertotal}{l}

%Component Index
\newcommand{\compidx}[1]{#1^{(\comp)}}

%MAC/pixel of a given module c
\newcommand{\macpx}{\compidx{m}_{\mathrm{px}}} 

%bias
\newcommand{\bias}{\mathrm{bias}}

%Learnable bitwidth (Originally b)
\newcommand{\bw}{b}
\newcommand{\bwreal}{\Tilde{b}}
\newcommand{\bwint}{\hat{\bw}}
\newcommand{\bwmax}{B^{+}} %(Originally b_max)
\newcommand{\bwmin}{B^{-}} %(Originally b_min)

%Learnable clipping bound
\newcommand{\clipbound}{v_{max}}

%Quantization bounds
% \newcommand{\qbmax}{Q_{max}} %(Originally Q_max) 
\newcommand{\qbmax}{Q^{+}} %(Originally Q_max) 
\newcommand{\qbmin}{Q^{-}} %(Originally Q_min)

% Quantization Index 
\newcommand{\qidxreal}{u} % Real-valued (Originally u) 
\newcommand{\qidx}{q} % Integer-valued (Originally q) 

%Quantization Scale (Originally s)
\newcommand{\scale}{s}

%Clipping bound of v (Originally v_max)
\newcommand{\vmax}{v^{+}}
\newcommand{\vmin}{v^{-}}
\newcommand{\wmax}[1][]{w^{+#1}}
\newcommand{\amax}[1][]{a^{+#1}}
\newcommand{\xmax}{x^{+}}
% \newcommand{\amax}[1]{a^{+#1}}
% \newcommand{\amax}{\overset{\scriptsize{+}}{a}}

%Scaling Factor (Originally SF)
\newcommand{\SF}{F}

%N for Scaling Factor (Originally SF)
\newcommand{\SFn}{n}

%PRECISIONS
%Variable has fp32 precision
\newcommand{\fp}[1]{#1^{\mathrm{fp32}}}

%Variable has int32 precision
\newcommand{\intw}[1]{\bar{#1}}

%Variable has int64 precision
% \newcommand{\intd}[1]{#1^{\mathrm{int64}}}
\newcommand{\intd}[1]{\bar{\bar{#1}}}

%Variable has int16 precision
\newcommand{\inth}[1]{#1^{\mathrm{int16}}}

\newcommand{\xfp}{\fp{x}}

%FUNCTIONS

\newcommand{\intf}[2]{\mathcal{I}_{#2}\left( #1 \right)}

\newcommand{\conv}{\mathrm{IntConv}}
\newcommand{\convf}[1]{\conv \left(#1\right)}

\newcommand{\clip}{\mathrm{clip}}
\newcommand{\clipf}[1]{\clip \left( #1 \right)}

\newcommand{\round}[1]{\left \lfloor#1 \right \rceil}

% quantized-and-reconstructed value
\newcommand{\quantf}[3]{\mathcal{Q}(#1;#2,#3)}
\begin{abstract}
Variational autoencoder-based neural video coding has demonstrated impressive rate-distortion performance. However, its adoption in real-world applications remains hindered by challenges, such as prohibitively high computational complexity and limited cross-platform interoperability. These issues are often overlooked, as most neural video codecs rely on floating‑point arithmetic to fully explore their rate-distortion potential. Practical deployment, however, requires integer-based implementations. Converting floating-point implementations into integer-based networks is non-trivial, since it involves quantizing inter-dependent coding components, whose sensitivity to precision may vary across codec designs. 
This paper introduces a Jointly-Optimized Mixed-Precision (JOMP) framework, in which both quantization parameters and bit widths are treated as learnable variables during training. This enables different codec modules to operate at varying precision levels, thereby jointly optimizing the rate-distortion-complexity trade-off. To the best of our knowledge, JOMP is the first mixed-precision quantization framework for neural video codecs. Its effectiveness is validated through a systematic investigation of quantization across different coding frameworks and temporal buffering strategies. Our study marks the first attempt to a unified understanding of the combined effects of modern coding frameworks and temporal buffering strategies, with the aim of informing future development of neural video codecs from a practicality perspective. In addition, we develop a complete integerization pipeline to achieve deterministic decoding.
Overall, when applied to our best-performing model, JOMP enables end-to-end mixed-precision learning for integer neural video codecs, achieving rate-distortion performance comparable to that of the state-of-the-art DCVC-FM while reducing bit operations by 87.6\%.
\end{abstract}

\section{Introduction}
\label{sec:intro}

\begin{figure}[t!]
    \begin{center}
        \centering
        \includegraphics[width=0.9\linewidth]{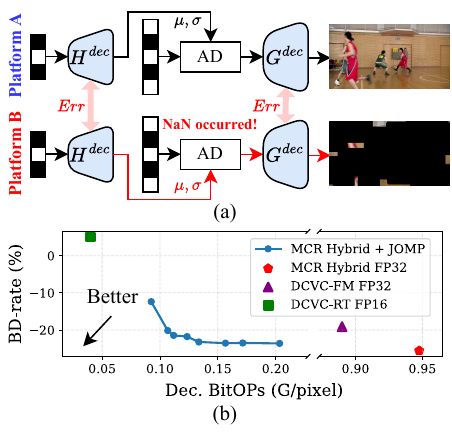}
        \vspace{-1.5 em}
        \caption{Motivation and overview of our work. (a) Floating-point decoding may suffer from cross-platform numerical inconsistency, where small differences can lead to inaccurate probability estimation and eventually decoding failure. (b) Jointly-Optimized Mixed-Precision (JOMP) quantization significantly reduces decoding complexity (BitOPs) while maintaining comparable BD-rate performance, achieving a favorable trade-off between coding efficiency and computational cost.}
        \label{fig:teaser}
    \end{center}
    \vspace{-1.5 em}
\end{figure}

\IEEEPARstart{C}{urrent} variational autoencoder-based neural video coding (NVC)~\cite{dcvc, canfvc, tcm, hem, dcvcdc, dcvcfm, dcvcrt, maskcrt, HyTIP} has been demonstrating superior rate-distortion efficiency in comparison to traditional codecs, highlighting its potential as a next-generation video coding paradigm. However, they typically rely on floating-point arithmetic, which hinders deployment in real-world scenarios. Due to the round-off error, floating-point computations may produce slightly different decoder-side outputs across devices that affect probability estimation used by entropy coding in NVC, resulting in decoding failure, as illustrated in Fig.~\ref{fig:teaser}(a). Furthermore, floating-point arithmetic generally consumes more energy than fixed-point arithmetic at comparable bit width, making them less attractive from an implementation perspective~\cite{FLIQS,tenlessons,Horowitz2014}. These limitations have motivated recent efforts towards integer/fixed-point NVC decoders as a means to mitigate cross-platform reconstruction mismatch~\cite{dcvcrt,mobilenvc}. This emerging direction broadens the design objectives of NVC beyond rate-distortion optimization by incorporating practical concerns, such as reproducibility, energy efficiency, computational complexity, and storage footprint. 

Despite this motivation, integerizing NVCs remains a non-trivial problem. For example, prior work~\cite{mobilenvc,dcvcrt} apply uniform-precision quantization schemes to a specific NVC model, making it unclear which codec components can be aggressively quantized and which require higher precision to preserve coding efficiency. Moreover, since these studies are limited to a single codec design, the extent to which their conclusions generalize across different coding frameworks remains underexplored.

Motivated by these gaps, this paper aims to advance integerized NVC design beyond prior works with a focus on key aspects that are critical for practical deployment. Specifically, we jointly optimize rate-distortion performance and computational complexity through a Jointly-Optimized Mixed-Precision (JOMP) quantization-aware training (QAT) framework. We use the number of bit operations (BitOPs)~\cite{mp-ptq, fracbits} as an indicator of platform-agnostic intrinsic computational complexity\footnote{Here, intrinsic computational complexity refers to the number of bit operations that is inherent to the algorithm and is independent of the underlying software or hardware platform.}. Since a video is typically encoded once but decoded many times, and because cross-platform inconsistencies arise from mismatches in the decoder path, we apply JOMP to the decoder side only. This choice directly targets the most critical stage for practical deployment, where deterministic reconstructions and reduced complexity are essential. Still, the proposed JOMP framework is not restricted to decoder-side optimization and could be applied end-to-end to the entire codec.

Rather than applying uniform bit width to the entire decoder, the JOMP framework accounts for how bit-precision reduction in each individual coding component affects both the overall platform-agnostic intrinsic complexity and the resulting rate-distortion performance. This enables the development of mixed-precision integer NVCs that better balances coding efficiency and complexity, as illustrated in Fig.~\ref{fig:teaser}(b). The resulting codec achieves rate-distortion performance comparable to DCVC-FM~\cite{dcvcfm} while offering substantially lower complexity. 

It is worth emphasizing that the proposed JOMP generalizes across distinct coding frameworks and temporal buffering strategies. This is particularly important because the impact of precision reduction depends strongly on the underlying codec design, since different coding schemes and buffering mechanisms induce varying trade-offs among reconstruction quality, temporal robustness, and computational complexity. This broad applicability across distinct NVC enables a systematic investigation across multiple coding frameworks and buffering strategies, allowing us to identify how precision sensitivity varies across diverse NVC design choices. Such an analysis provides insights that extend beyond a single codec configuration and is therefore essential for advancing the development of practical and deployable integer NVCs.

Our main contributions are summarized as follows:
\begin{itemize}
\item We develop a Jointly-Optimized Mixed-Precision (JOMP) quantization-aware training framework for neural video codecs (NVCs), enabling joint optimization of coding efficiency and computational complexity in an end-to-end manner.

\item We present the first systematic study of integerized neural video codecs across multiple coding frameworks and temporal buffering strategies, providing new insights into precision sensitivity and their rate-distortion-complexity trade-offs. This broad applicability is enabled by the generality of JOMP.

\item We design a complete integerization pipeline for neural video codecs, including integer convolution and integer implementations of key modules, enabling bit-exact and cross-platform consistent decoding.

\item We demonstrate that the proposed framework achieves competitive coding performance while significantly reducing decoding complexity, contributing toward the practicality of neural video codec deployment.

\end{itemize}

\section{Background} 
This section addresses the fundamentals of neural video coding and neural network quantization, offering the necessary background to facilitate comprehension of our work.

\subsection{Neural Video Coding} \label{sec:background_nvc}
\begin{figure}[t!]
    \begin{center}
    \includegraphics[width=\linewidth, trim=0.5cm 0cm 1cm 0cm,clip]{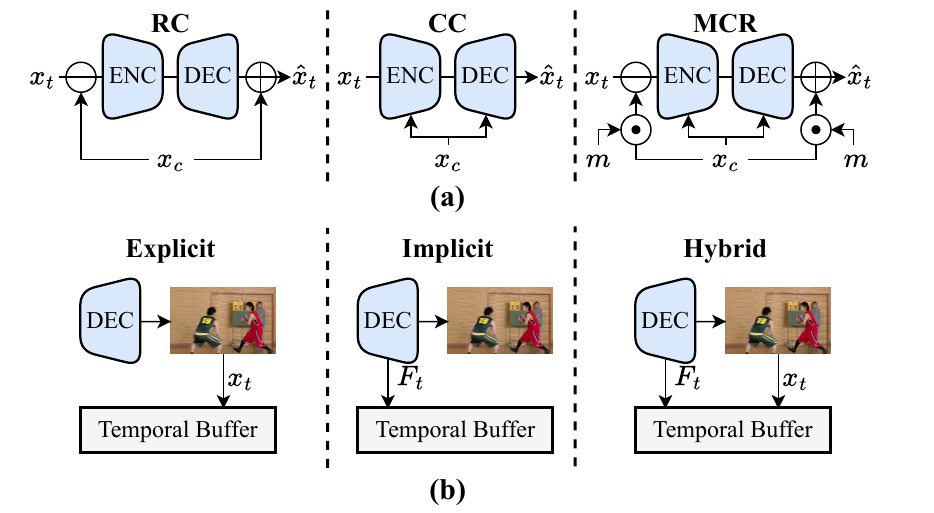}
    \vspace{-2 em}
    \caption{Overview of NVC design variants considered in this work. (a) Different coding frameworks for inter-frame compression. (b) Different temporal buffering strategies used to propagate past information across frames.}
    \label{fig:variants}
    \end{center}
    \vspace{-1.5em}
\end{figure}

Modern NVCs vary substantially in both their coding frameworks and temporal buffering strategies. Fig.~\ref{fig:variants} provides an overview of these two design dimensions, where Fig.~\ref{fig:variants}(a) illustrates different coding frameworks and Fig.~\ref{fig:variants}(b) presents different temporal buffering strategies. 
In the figure, $x_t$ denotes the current frame to be compressed, and $\hat{x}_t$ represents its reconstructed version at the decoder. The variable $x_c$ is the motion-compensated reference (or its corresponding features), which serves as a prediction signal to facilitate compression of the current frame. In masked conditional residual coding (MCR), a spatial mask $m$ is further introduced to modulate the contribution of the reference. Additionally, $f_t$ denotes the learned feature representation of the current frame, which can be stored in a temporal buffer to assist the decoding of subsequent frames.

\subsubsection{Coding frameworks}
Existing inter-frame coding frameworks can be broadly categorized into three types: \emph{Residual Coding }(RC), \emph{Conditional Coding }(CC), and \emph{Masked Conditional Residual Coding }(MCR) (see Fig.~\ref{fig:variants}(a)).

\textbf{Residual Coding (RC)}~\cite{dvc, mlvc, nvc, ssf, fvc} directly encodes the differences $x_t - x_c$ between the current frame $x_t$ and its motion-compensated prediction $x_c$, following the classical paradigm established in traditional video codecs. Despite its conceptual simplicity, RC provides a strong and interpretable baseline.

\textbf{Conditional Coding (CC)}~\cite{dcvc, canfvc, tcm, hem, dcvcdc, dcvcfm, dcvcrt} departs from the residual paradigm by treating the predicted frame or its extracted features as conditioning signals that jointly guide both the encoder and decoder. This formulation enables richer temporal modeling and has become the dominant approach in recent high-performance neural video codecs.

\textbf{Masked Conditional Residual Coding (MCR)}~\cite{maskcrt, HyTIP}, inspired by Brand {\it et al.}~\cite{crc}, further extends CC by introducing a spatial mask $m$ that adaptively subtracts the reference prediction $x_c$ from the original frame $x_t$ at the sub-frame level before passing the resulting signal $x_t - m \odot x_c$ through a conditional inter-frame codec. This mechanism helps alleviate the information bottleneck~\cite{crc} in CC and improves coding efficiency.

\subsubsection{Temporal buffering strategies}
To propagate temporal information across frames, neural video codecs maintain intermediate states in a temporal buffer. Existing approaches can be grouped into three types~\cite{HyTIP} (see Fig.~\ref{fig:variants}(b)).

\textbf{Explicit buffering}~\cite{dvc, mlvc, nvc, ssf, fvc, dcvc, canfvc, maskcrt} stores previously reconstructed frames and reuses them as references for future frames. This approach aligns closely with traditional codecs and offers strong interpretability, but may propagate reconstruction artifacts over time.

\textbf{Implicit buffering}~\cite{tcm, hem, dcvcdc, dcvcfm, dcvcrt} propagates learned latent features across frames, bypassing the need to store reconstructed frames. While this enables expressive temporal modeling, it introduces greater sensitivity to quantization errors, as learned features are often not regularized to be compact and robust to quantization.

\textbf{Hybrid buffering}~\cite{HyTIP} combines reconstructed frames with learned latent features, aiming to leverage the complementary strengths of both strategies. This design has demonstrated strong empirical performance.

\subsubsection{Model variants} 
\label{sec:model_variants}
Coding frameworks and temporal buffering strategies are treated as orthogonal design dimensions in this work. Each of the three coding frameworks can, in principle, be paired with any of the three buffering strategies, yielding nine distinct model variants. We systematically analyze all nine configurations to understand how quantization affects rate-distortion-complexity trade-offs across diverse codec designs.

\subsection{Neural Network Quantization} 
\label{sec:quantization}
Quantization is the process of mapping a large, potentially infinite, uncountable set of continuous values to a finite, countable set of discrete values. In the context of neural networks, it has become an important technique because it reduces memory footprint and computational cost by representing weights, activations, and other model parameters using low-precision formats. As a result, quantized models can achieve smaller model sizes, faster inference, and lower energy consumption, all of which are critical in latency-sensitive and resource-constrained applications. 

Quantization can be categorized in several ways based on: (i) the spacing of its quantization levels (uniform or non-uniform), (ii) how the input data are processed (scalar or vector quantization), and (iii) whether real-valued inputs are mapped to discrete integer values centered around zero (symmetric) or with an offset (asymmetric). This work, as the following discussion, focuses on \textit{symmetric, uniform, and scalar} quantization.

Given a real-valued input $v \in \mathbb{R}$, quantization maps it to the quantization-index domain $u$ via a scale factor $\scale$, as demonstrated in~\eqref{eq:quant_real_u}. This value is then clipped to the range determined by the bit width $b$ and rounded to obtain an integer index. We formalize this operation as an integerization function $\intf{\cdot}{b}$ in~\eqref{eq:quant_intf}, where $\qbmin_{b}=-2^{b-1}$ and $\qbmax_{b}=2^{b-1}-1$. Accordingly, the quantization index $\bar v^\qidx_{b}$ of $v$ is given in~\eqref{eq:quant_idx}. From the scale $\scale$ and quantization index $\bar v^\qidx_{b}$, one can obtain the reconstructed $\hat v$ value, as shown in~\eqref{eq:dequant}.
\begin{equation}
    \label{eq:quant_real_u}
    \qidxreal = \frac{v}{\scale}
\end{equation}
\begin{equation}
    \label{eq:quant_intf}
    \intf{u}{\bw} \triangleq \round{\clip(u,\,\qbmin_{b},\,\qbmax_{b})}
\end{equation}
\begin{equation}
    \label{eq:quant_idx}
    \bar v^\qidx_{b} = \intf{\qidxreal}{b}
\end{equation}
\begin{equation}
    \label{eq:dequant}    
    \hat{v} = \bar v^\qidx_{b} \times \scale
\end{equation}

From~\eqref{eq:quant_real_u}, \eqref{eq:quant_intf}, \eqref{eq:quant_idx}, and \eqref{eq:dequant}, the reconstructed value $\hat v$ can be directly obtained from $v$ once $\scale,b$ are defined. For this, we define the quantization-dequantization function $\mathcal{Q}(\cdot)$ in~\eqref{eq:quant_dequant}, which is commonly referred to as fake (or simulated) quantization because both $\scale$ and $\hat v$ remain floating-point values. In practical fixed-point implementations, however, both must be represented in integer form.
\begin{equation}
    \label{eq:quant_dequant}
    \quantf{v}{\scale}{b} \triangleq \intf{v/\scale}{b} \times \scale = \hat v
\end{equation}

The scale factor $\scale$ can be defined by a clipping bound $\vmax$ as in~\eqref{eq:cal_scale}, where $\vmax$ controls the representable dynamic range. Notably, this quantity is not necessarily equal to the maximum value attained by $v$, since outlier values may be intentionally clipped. 

\begin{equation}
    \label{eq:cal_scale}
    \scale = \frac{\vmax}{\qbmax_{b}},
\end{equation}

\begin{figure}[t!]
    \begin{center}
    \includegraphics[width=\linewidth, trim=0cm 0cm 0cm 0.5cm,clip]{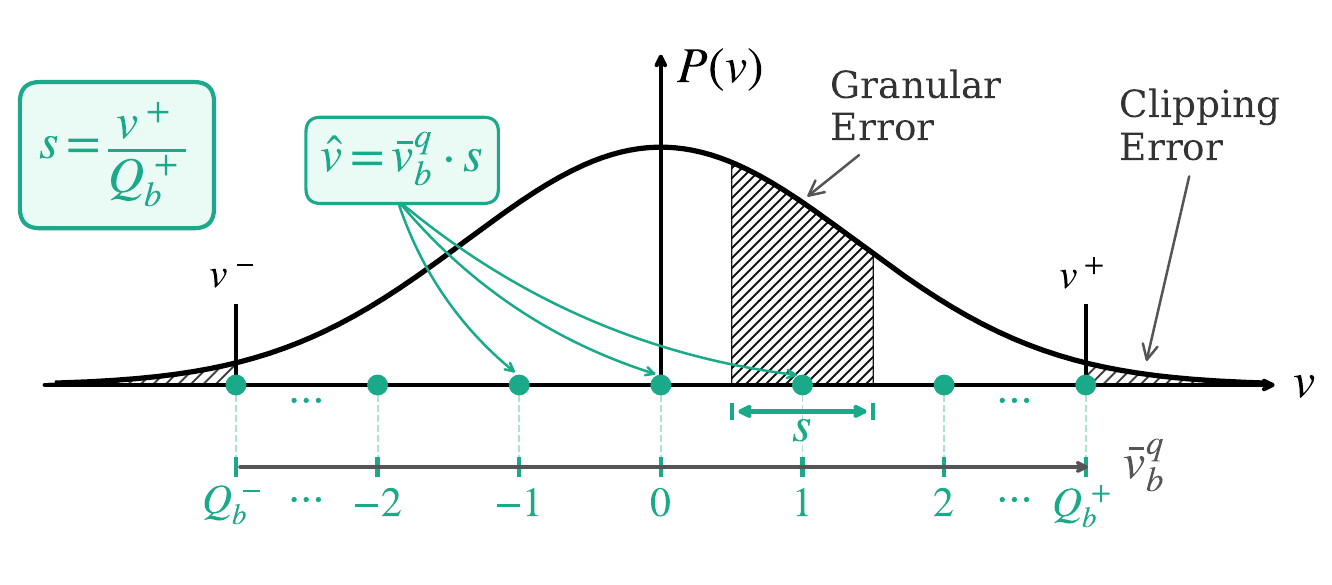}
    \vspace{-3 em}
    \caption{Illustration of a symmetric, uniform, and scalar quantization scheme along with a data distribution $P(v)$. The input value $v$ is quantized to the index $\bar{v}^{q}_{b}$ using the step size $s$, with clipping bounds $\vmin,\vmax$ and corresponding quantization indices $\qbmin_b,\qbmax_b$. The reconstructed value is denoted by $\hat{v}$. Quantization introduces granular errors for values within the clipping range and clipping errors for values outside.}
    \label{fig:quantization_diagram}
    \end{center}
    \vspace{-1.5em}
\end{figure}

It can be observed that learning any two among $\vmax$, $\scale$, and $\qbmax_{b}$ automatically determines the third. Thus, for a fixed bit width, $\scale$ governs the granular quantization error, which arises from the finite spacing between adjacent quantization levels; whereas $\vmax$ controls the clipping error, which arises when values exceed the representable range and are truncated to the nearest boundary. Their relationship is illustrated in  Fig.~\ref{fig:quantization_diagram}. In conventional fixed-bit-width quantization approaches, as discussed next, the quantization process must define either $\vmax$ or $\scale$, with the other being inferred accordingly, thereby determining the trade-off between granular and clipping errors. Nevertheless, this process may still rely on a bit-width choice that could be itself overestimated or underestimated.

Quantization in neural networks is commonly realized through two main approaches: \emph{Post-Training Quantization} (PTQ) and \emph{Quantization-Aware Training} (QAT)~\cite{aimet}, in both of which the bit width $b$ (or, equivalently, $\qbmax_{b}$) is treated as a predefined hyperparameter. As the name suggests, PTQ determines the quantization scale factor $\scale_v$ after training by minimizing the mean-squared reconstruction error, without modifying the model parameters. Equation~\eqref{eq:ptqopt} illustrates this process, where $v$ denotes either activations or network weights, and $b_v$ is the predefined bit width assigned to $v$. For activations, a small calibration dataset is typically used to evaluate the mean-squared reconstruction error and estimate $\scale_v$.
\begin{equation}
    \label{eq:ptqopt}
    \scale_v^{*}
    =
    \arg\min_{\scale_v}
    \mathrm{MSE}\!\left(v,\quantf{v}{\scale_v}{b_v}\right),
\end{equation}

In contrast, QAT incorporates quantization effects during training by jointly optimizing the model parameters $\theta$ together with the quantization scale factors $\scale_a$ and $\scale_w$ for activations and weights, respectively. Specifically, training is performed using the fake-quantized activation $\hat{a}=\quantf{a}{\scale_a}{b_a}$, the fake-quantized weight $\hat{w}=\quantf{w}{\scale_w}{b_w}$, and the ground-truth target $y$, as demonstrated in~\eqref{eq:quant_theta} and~\eqref{eq:qatopt}. The fake-quantized activations $\hat{a}$ and weights $\hat{w}$ assume discrete real-valued levels, although they are still represented in floating-point precision. In~\eqref{eq:qatopt}, $\mathcal{M}$ may represent an entire model, an individual layer, or a component of an NVC, while $\mathcal{J}$ denotes the training objective. The bias term in~\eqref{eq:quant_theta} is typically retained in high or full precision and is therefore not associated with quantization parameters.

\begin{equation}
    \label{eq:quant_theta}
    \hat{\theta}=\{\hat{w},\bias\},
\end{equation}
\begin{equation}
    \label{eq:qatopt}
    (\hat{\theta}^{*},\scale_a^{*},\scale_w^{*})
    =
    \arg\min_{\hat{\theta},\scale_a,\scale_w}
    \mathcal{J}\!\left(y,\mathcal{M}(\hat{a};\hat{\theta})\right).
\end{equation}

Although QAT incorporates quantization effects during optimization, both PTQ and QAT typically assume fixed bit widths \(b_a\) and \(b_w\). As a result, \(\qbmax_b\) is predefined rather than learned, and the learned scale factor \(\scale\) directly determines the clipping bound \(\vmax\), as shown in~\eqref{eq:cal_scale}. This reduction in the degrees of freedom from two to one introduces an undesirable trade-off between granular and clipping errors, as previously discussed.

\section{Related Works}
\label{sec:related_work}

This section reviews prior work on the quantization of neural image and video compression models, and also discusses previous studies focused on mixed-precision quantization strategies for neural networks in a broader context.

\subsection{Quantized Neural Image and Video Codecs}
\begin{table}[t]
\centering

\centering
\renewcommand{\arraystretch}{1.0}
\caption{Comparison with quantization methods in neural image codecs.}
\begin{tabular}{l|c|c|c}
\hline
Methods & Precision & \begin{tabular}[c]{@{}c@{}}Bit-width\\ Assignment\end{tabular} & \begin{tabular}[c]{@{}c@{}}Learnable\\ Bit-widths\end{tabular} \\
\hline

JPEG AI~\cite{jpegai}
& W8A8
& Manual
& $\mathbf{\times}$ \\

FPX-NIC~\cite{fpxnic}
& W8A8
& Manual
& $\mathbf{\times}$ \\

Koyuncu {\it et al.}~\cite{cross_device} 
& W16A16 
& Manual 
& $\mathbf{\times}$ \\

% CW-NLQ~\cite{fixedquant} 
% & W8 
% & Manual 
% & $\mathbf{\times}$ \\

Q-LIC~\cite{qlic}
& W8A8 
& Manual 
& $\mathbf{\times}$ \\

MP-PTQ~\cite{mp-ptq} 
& MP 
& Search-based 
& $\mathbf{\times}$ \\

FMPQ~\cite{immixed} 
& MP 
& Search-based 
& $\mathbf{\times}$ \\

\hline
Ours 
& MP 
& Optimization-based 
& \checkmark \\
\hline
\end{tabular}
\label{tab:comparison_qnic}
\vspace{-1em}
\end{table}

\begin{table}[t]
\centering
\renewcommand{\arraystretch}{1.0}
\caption{Comparison with quantization methods in neural video codecs.}
\begin{tabular}{l|c|c|c}
\hline
Methods & Precision & Frameworks & Buffer \\
\hline
MobileCodec~\cite{mobilecodec} 
& W8A8 
& RC 
& Explicit \\

MobileNVC~\cite{mobilenvc} 
& W8A8 
& RC 
& Explicit \\

DCVC-RT~\cite{dcvcrt}   
& W16A16 
& CC 
& Implicit \\

\hline
Ours      
& MP 
& RC/CC/MCR 
& Exp./Imp./Hybrid \\
\hline
\end{tabular}
\label{tab:comparison_qnvc}
\vspace{-1em}
\end{table}
Regarding variational autoencoder-based neural image codecs, the prior work~\cite{cross_device, jpegai} shows that quantizing weights and activations in hyperprior components can mitigate cross-platform inconsistencies, while~\cite{qlic, balle_integer, jpegai_deterministic} apply quantization to the entire decoder. Beyond these approaches, 
%more advanced strategies, such as group-wise quantization~\cite{fixedquant} and
mixed-precision methods~\cite{mp-ptq,immixed} have been proposed to better balance compression performance and computational complexity. 
Toward practical deployment,~\cite{fpxnic} further demonstrates an FPGA implementation of a neural image codec based on uniform symmetric quantization.
As summarized in Table~\ref{tab:comparison_qnic}, most existing approaches for neural image codecs either rely on fixed-precision quantization (e.g., W8A8, where W and A denote the bit widths of weights and activations, respectively) or adopt mixed-precision (MP) quantization using search-based strategies to assign bit widths for various components.

In contrast, quantization in variational autoencoder-based NVCs remains relatively underexplored. As summarized in Table~\ref{tab:comparison_qnvc}, existing approaches mostly adopt uniform precision across components or apply quantization in a limited scope. For example, MobileCodec~\cite{mobilecodec} and MobileNVC~\cite{mobilenvc} employ W8A8 quantization using a PTQ+QAT pipeline, while DCVC-RT~\cite{dcvcrt} adopts W16A16 quantization to ensure deterministic inference. Other works investigate partial quantization, such as applying 8-bit quantization to the hyperprior decoder~\cite{ruhan}. However, these methods do not consider the heterogeneous nature of NVC components, where different components may exhibit distinct sensitivity to quantization.

Quantization has also been explored in implicit neural representation-based image and video codecs~\cite{c3,hinerv}. However, their primary objective is bitrate reduction through network weight compression, rather than deterministic decoding or cross-platform interoperability.

Motivated by these limitations in the literature on variational autoencoder-based neural video codecs, we investigate quantization in NVCs from a broader perspective through the proposed JOMP framework. JOMP adaptively allocates bit widths across codec components to achieve improved rate-distortion-complexity trade-offs, as presented in Sec.~\ref{sec:method}, with the goal of enabling deployment-oriented integer inference.

\subsection{Mixed-Precision Quantization}
In mixed-precision quantization, a key design problem is how to assign different bit widths to network components while balancing model performance and computational cost. Existing approaches can be broadly categorized into two classes: search-based and optimization-based methods.

In search-based methods, e.g., MP-PTQ~\cite{mp-ptq} and FMPQ~\cite{immixed}, bit-width assignments are typically determined through iterative evaluation of candidate bit-width configurations, where each bit-width configuration is trained or partially evaluated to estimate the resulting model performance and complexity cost. In these methods, the update of model weights is generally performed as a separate task and is independent of the bit-width assignment. 

In contrast, optimization-based methods treat bit widths as learnable parameters and jointly optimize  bit-width assignments together with network weights during training, enabling a more direct and fine-grained exploration of the performance-complexity trade-off. Because the loss function is non-differentiable with respect to integer bit widths, existing optimization-based approaches address this challenge through various relaxations. Uhlich {\it et al.}~\cite{uhlich2020} enable gradient-based optimization by parameterizing quantization step sizes and clipping bounds as optimization variables. FracBits~\cite{fracbits} introduces fractional bit widths with linear interpolation, while SDQ~\cite{SDQ} models bit-width allocation as a stochastic process using differentiable sampling (e.g., Gumbel-softmax). DDQ~\cite{DDQ} further proposes a differentiable formulation by representing quantization as a learnable matrix-vector operation. 

\section{Jointly-Optimized Mixed-Precision QAT}
\label{sec:method}

\begin{figure}[t!]
    \begin{center}
    \includegraphics[width=0.9\linewidth]{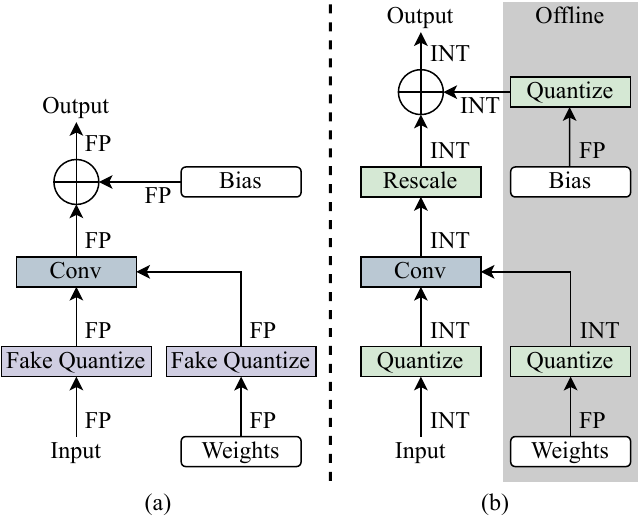}
    \vspace{-1 em}
    \caption{Comparison between fake-quantization and fully integerized convolution. (a) Fake quantization emulates low precision while computations remain in floating point. (b) Fully integerized convolution performs all operations in integer arithmetic. 
    }
    \label{fig:sim_real}
    \end{center}
    \vspace{-1.5em}
\end{figure}

\begin{figure*}[t!]
    \begin{center}
    \includegraphics[width=0.9\linewidth]{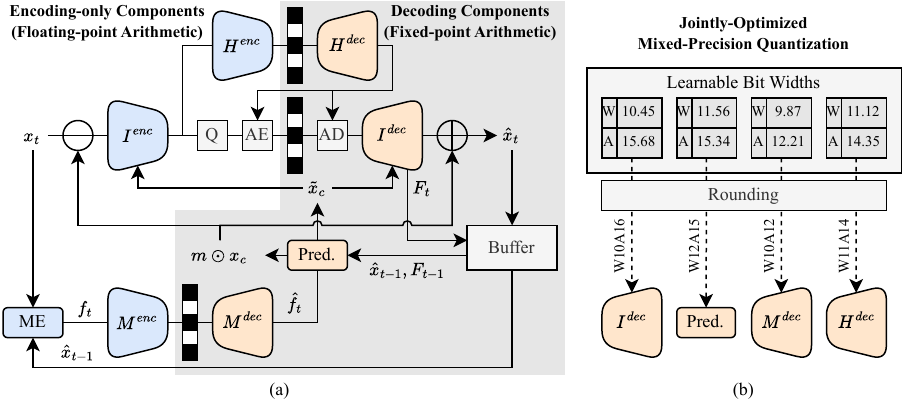}
    \vspace{-1 em}
    \caption{Overview of the quantization framework. (a) Decoder-side quantization architecture based on the MCR Hybrid framework. We apply quantization only to the decoder and decompose it into four components: the inter-frame main decoder $I^{dec}$, the prediction network $\mathrm{Pred.}$, the motion decoder $M^{dec}$, and the hyperprior decoder $H^{dec}$. (b) Jointly-Optimized Mixed-Precision quantization. Bit widths are treated as learnable parameters and jointly optimized during training, enabling automatic bit-width assignment for different decoder components.}
    \label{fig:high-level-architecture}
    \end{center}
    \vspace{-1.5em}
\end{figure*}

NVC integerization process used in this work consists of two stages: (i) Jointly-Optimized Mixed-Precision (JOMP) quantization-aware training (QAT), in which the clipping bounds, component-wise bit widths, and network weights are jointly optimized through fake-quantization (see Fig.~\ref{fig:sim_real}(a)), as described in this section; and (ii) the conversion of the fake-quantized NVC into an actual integer NVC (see Fig.~\ref{fig:sim_real}(b)), as described in Sec.~\ref{sec:integer_network}, enabling deterministic and consistent deployment across heterogeneous platforms.
The same framework is applied to the nine variants described in Sec.~\ref{sec:background_nvc}, spanning three coding frameworks and three temporal buffering strategies.

As shown in Fig.~\ref{fig:high-level-architecture}(a), NVC consists of four main components: 
(i) the \textbf{inter-frame main codec} ($I_{\text{enc}}$/$I_{\text{dec}}$), 
responsible for encoding and decoding the target frame based on predicted signals; 
(ii) the \textbf{hyperprior codec} ($H_{\text{enc}}$/$H_{\text{dec}}$), 
which models the latent distribution for entropy coding; 
(iii) the \textbf{motion codec} ($M_{\text{enc}}$/$M_{\text{dec}}$), 
which compresses the motion information between frames; and 
(iv) the \textbf{prediction network} (Pred.), 
which generates the predicted signals based on temporal context and motion information.

Our quantization and integerization processes focus exclusively on the \textit{decoding} components (Fig.~\ref{fig:high-level-architecture}(a)), where bit-exact reconstruction across different deployment platforms is essential to prevent temporal drift. Reducing the complexity of these decoding components is crucial for practical deployment, since decoding is often executed repeatedly and efficiently across diverse devices under constraints of latency, energy consumption, and memory footprint. By contrast, encoding-only components remain in full floating-point precision, since they do not affect cross-platform consistency and a video is generally encoded once but decoded multiple times.

As discussed in Sec.~\ref{sec:quantization}, conventional QAT treats bit widths as predefined design choices rather than learnable parameters. In the context of NVC, which consists of multiple deep neural network components, QAT typically leads to either uniform precision across all components or empirically-designed mixed-precision assignments.
To overcome this limitation, our JOMP framework jointly optimizes not only the scale factors (or clipping bounds) and network weights, as in conventional QAT, but also the bit width associated with each component. However, without additional constraints in the loss function, such optimization would naturally favor higher quantization precision, since larger bit widths reduce quantization errors. This motivates the use of a complexity-aware regularization term, enabling learnable precision while avoiding trivial high-precision solutions.

In our design, all layers within a given NVC component share the same bit width, whereas the corresponding quantization clipping bounds are learned independently for each convolutional layer to better adapt to their distinct activation and weight dynamic ranges. In this way, JOMP establishes a component-wise mixed-precision formulation for integerized NVCs, where precision assignment is optimized jointly with coding performance and complexity. 
The component-wise bit-width constraint may allow the re-use of the same hardware adders and multipliers across convolutional layers in the same component. However, it is important to mention that this constraint can be adjusted to better accommodate specific application requirements.

It is worth mentioning that we adopt a uniform, symmetric quantization framework due to its hardware efficiency and consistency across modules. From a system perspective, uniform quantization enables efficient integer arithmetic with a single scale factor, which simplifies deployment and reduces implementation overhead. The symmetric design~\cite{cross_device,mp-ptq,fpxnic} further eliminates zero-point offsets, leading to more efficient computation and easier accumulation in integer pipelines.

\subsection{JOMP Loss Function}

\label{sec:jomp}
To formalize the trade-off between coding efficiency and complexity, we start from the conventional rate-distortion objective described in~\eqref{eq:rd_loss}, where $R$ and $D$ denote the bitrate and distortion terms, respectively, and $\lambda_\text{D}$ controls the rate-distortion trade-off. Building upon this formulation, we further incorporate computational complexity into the training objective through a BitOPs regularization term, as demonstrated in~\eqref{eq:full_loss}, where $\mathcal{L}_{\text{BitOPs}}$ directly reflects decoding complexity in a normalized form, while  $\lambda_{\text{C}}$ controls the trade-off between coding performance and computational cost.
The complexity loss $\mathcal{L}_{\text{BitOPs}}$ is defined in~\eqref{eq:bop_loss}, where $\macpx$ denotes the number of per-pixel multiply-accumulate operations associated with component $\comp$. The variables $\compidx{\bw}_w$ and $\compidx{\bw}_a$ denote the corresponding weight and activation bit widths of component $\comp$, respectively. Since all layers within a given component share the same bit width, complexity is assessed at the component level. The multiplication by $10^{-9}$ converts the BitOPs count per pixel into giga-BitOPs per pixel (G/pixel).
\begin{equation}
    \mathcal{L}_{\text{RD}} = R + \lambda_\text{D} D
    \label{eq:rd_loss}
\end{equation}
\begin{equation}
    \label{eq:full_loss}
        \mathcal{L}_{\text{JOMP}} = \mathcal{L}_{\text{RD}} + \lambda_{\text{C}} \mathcal{L}_{\text{BitOPs}}
\end{equation}
\begin{equation}
    \label{eq:bop_loss}
    \mathcal{L}_{\text{BitOPs}} = 10^{-9}\sum_{\comp} \macpx\, \compidx{b}_w\, \compidx{b}_a
\end{equation}

\subsection{Learning Quantization Parameters and Bit Widths}
\label{sec:learn_quant_param_and_bitwidth}
In this subsection we adopt the same notation introduced in Sec.~\ref{sec:quantization} to derive the gradients of the loss function in~\eqref{eq:full_loss} with respect to quantization parameters for gradient-based optimization.%Additionally, the formulation we use is derived from prior works~\cite{liang2025,uhlich2020} which enable gradient-based optimization of quantization parameters.

\subsubsection{Problem Formulation} \label{sec:problem_form}

Let $v \in \mathbb{R}$ denote a real-valued input to be quantized, which, in our case, corresponds to a model weight or an activation. The goal is to learn both the clipping bound $\vmax \in [0, +\infty)$ and the continuous bit-width parameter $\bwreal \in [8, 16]$.
Notably, from the clipping bound and bit width, the quantization scale $s$ can be derived according to~\eqref{eq:cal_scale}. Since non-integer bit widths are not valid in practice, we project the continuous learnable bit-width $\bwreal$ onto a discrete finite interval:
% Both parameters are jointly optimized during training to minimize the overall loss function described in~\eqref{eq:full_loss}. Note that, from learned $\vmax$ and the projected bitwidth $\bw$, the quantization parameters can be calculated according to~\eqref{eq:quant:scaler}.
\begin{equation}
    \bw = \round {\clip(\bwreal, \bwmin, \bwmax)},
    \label{eq:quantize_b}
\end{equation}
where we set $\bwmin=8$ and $\bwmax=16$ in our implementation. Therefore, the actual bit width $b$ assigned to any module is an integer from 8 to 16 inclusive.

As shown in Fig.~\ref{fig:high-level-architecture}(b), assuming that a given NVC decoder has $\comptotal$ components, we define the set of bit widths for weights and activations as
\begin{equation}
    \label{eq:bidwidths}
    \beta \triangleq \left\{
        \left( \bwreal_w^{(1)}, \bwreal_a^{(1)} \right),
        \cdots,
        \left( \bwreal_w^{(\comptotal)}, \bwreal_a^{(\comptotal)} \right)
    \right\}.
\end{equation}

Additionally, let $\layertotal$ denote the total number of layers in the decoder across all components of the model. Since each layer has its own pair of clipping bounds, or equivalently, its own pair of quantization scales for weights and activations, we define the set $V$ of all clipping bounds as
\begin{equation}
    \label{eq:clippingbounds}
    V \triangleq \left\{
        \left( w^{+[1]}, a^{+[1]} \right),
        \cdots,
        \left( w^{+[\layertotal]}, a^{+[\layertotal]} \right)
    \right\}.
\end{equation}

Therefore, the objective of JOMP is to minimize the loss function in~\eqref{eq:full_loss} through a gradient-descent optimization process, as formalized in~\eqref{eq:jompopt}.
\begin{equation}
    \label{eq:jompopt}
    (\hat \theta^{*},V^{*},\beta^{*})
    =
    \arg\min_{\hat \theta,V,\beta}
    \mathcal{L}_{\text{JOMP}}
\end{equation}

In terms of the quantizer design, the key distinction of JOMP from~\eqref{eq:qatopt}, which represents conventional QAT in its typical form, is that most QAT methods learn only quantization scales under fixed bit widths. According to~\eqref{eq:cal_scale}, once the bit width $b$ is specified and $\qbmax_{b}$ is determined, learning the quantization scale $s$ reduces to learning only the clipping bound $\vmax$. In contrast, JOMP jointly learns both bit widths and clipping bounds during optimization. As illustrated in Fig.~\ref{fig:quantization_diagram}, the clipping bounds govern clipping errors while the bit widths control granular errors within those clipping bounds. Consequently, with fixed bit widths in conventional QAT, one must trade off between these two types of errors, whereas JOMP mitigates both simultaneously under a complexity constraint.

\subsubsection{Gradient-Based Optimization} 

Building on the quantization formulation in Sec.~\ref{sec:quantization}, we derive the gradient expressions of the quantized value $\hat v$ with respect to the real-valued input $v$, the clipping bound $\vmax$, and the bit-width parameter $\bwreal$, as presented in~\eqref{eq:ste_vhat_v}, \eqref{eq:ste_vhat_vmax}, and \eqref{eq:ste_vhat_breal}, respectively. These gradients enable end-to-end optimization of the network weights together with $\vmax$ and $\bwreal$ via backpropagation. The resulting expressions are piecewise-defined depending on whether the value $u$ in~\eqref{eq:quant_real_u} lies inside or outside the clipping range in~\eqref{eq:quant_intf}. To handle the rounding operation in the backpropagation, we adopt the straight-through estimation (STE)~\cite{ste} approximation.  The detailed derivations are provided in the appendix.

\textbf{Gradient w.r.t. Input $v$:} 
\begin{equation}    
    \label{eq:ste_vhat_v}
    \frac{\partial \hat{v}}{\partial v} =
    \begin{cases}
    1, & \qbmin_{\bw} < u < \qbmax_{\bw} \\
    0, & \text{otherwise.}
    \end{cases}
\end{equation}
\textbf{Gradient w.r.t. Clipping Bound $\vmax$:}
\begin{equation}    
    \label{eq:ste_vhat_vmax}
    \frac{\partial \hat{v}}{\partial \vmax} =
    \begin{cases}
    \frac{\lfloor u \rceil - u}{\qbmax_{\bw}}, & \qbmin_{\bw} < u < \qbmax_{\bw} \\
    \frac{\qbmin_{\bw}}{\qbmax_{\bw}}, & u \le \qbmin_{\bw} \\
    1, & u \ge \qbmax_{\bw}.
    \end{cases}
\end{equation}
\textbf{Gradient w.r.t. Continuous Bit Width $\bwreal$:}
\begin{equation}    
    \label{eq:ste_vhat_breal}
    \frac{\partial \hat{v}}{\partial \bwreal} =
    \begin{cases}
    (v-\hat{v}) \times \frac{2^{\bwreal - 1}\ln 2}{\qbmax_{\bw}}, & \qbmin_{\bw} < u < \qbmax_{\bw} \\
    \vmax \times \frac{2^{\bwreal-1} \ln 2}{(\qbmax_b)^2}, & u \leq \qbmin_{\bw} \\
    0, & u \geq \qbmax_{\bw}.
    \end{cases}
\end{equation}

\section{Neural Video Compression Integerization} \label{sec:integer_network}

JOMP relies on fake-quantized convolutions during training to enable gradient-based optimization. However, for practical deployment, these simulated operations must be converted into true integer operations, since fake quantization is still performed using floating-point arithmetic. This conversion is particularly important because most NVC components are dominated by convolution operations. Moreover, convolution is the main operation from the perspectives of both computational complexity and cross-platform interoperability for NVCs.
As shown in Fig.~\ref{fig:sim_real}(b), all operations and operands are converted into integers at this stage, where no training or learning takes place except integer conversion.

Although floating‑point and integer convolution are conceptually similar, deploying integer convolution in practice poses additional challenges. Generally, activations and weights are quantized with different scales, namely $\scale_a$ and $\scale_w$. The key challenges are: (i) how to efficiently perform convolution when both activations and weights are represented in integer form but correspond to distinct quantization scales, and (ii) how to propagate results across convolutional layers with varying quantization parameters. To address these challenges, Sec.~\ref{sec:integer_conv} presents the Deployment-Aware Integer Convolution Block, which efficiently handles pre- and post-processing required for integer convolution. Sec.~\ref{sec:nonconv_integer} then describes the integerization procedure for the remaining non-convolutional layers.

\subsection{Deployment-Aware Integer Convolution Block} 
\label{sec:integer_conv}

\begin{algorithm}[t]
\caption{Deployment-Aware Integer Convolution Block}
\label{alg:integer_conv}
\begin{algorithmic}[1]
% \State \textbf{Input:} $\bar {x}$, $\bar{w}_{\bw_w}^q$, $\overline {\bias}$, $\bar{\scale}_{in}$, $\bar{\scale}_{out}$, $\SFn_a$, $\SFn_{\text{in}}$, and $\SFn_{\text{out}}$
\State \textbf{Input:} Input feature $\bar {x}$; Weight $\bar{w}_{\bw_w}^q$; Bias $\overline {\bias}$; Scale factors $\bar{\scale}_{in}$ and $\bar{\scale}_{out}$; Hyperparameters $\SFn_a$, $\SFn_{\text{in}}$, and $\SFn_{\text{out}}$.
\State \textbf{Output:} $\bar{y}$
\State $\bar{x}^{q}_{\bw_a} \gets \intf{(\bar {x} \times \bar{\scale}_{in}) \gg (\SFn_a + \SFn_{\text{in}})}{\bw_a}$ \Comment{Input quant.}
\State $\bar{\bar{a}}^{q} \gets \convf{\bar{x}^{q}_{\bw_a}, \bar{w}_{\bw_w}^q}$ \Comment{Integer convolution}
% \State $a^{int64} \gets a^{int64} \times s_{out}^{int32} / (SF_{out} / SF_{x})$ \Comment{Output scaling}
% \item[]
% \State $\intd{a}_{h}^{q} \gets \intd{a}^{q} \gg 32$ \Comment{Overflow-safe rescaling}
% \State $\intd{a}_{l}^{q} \gets \intd{a}^{q}\, \&\ \text{0xFFFFFFFF}$
% \State $\intd{a}_\text{aux} \gets \left(\intd{a}_{h}^{q} \times \bar{\scale}_{out} \right) \ll (32 - \left(\SFn_{\text{out}} - \SFn_{a} \right))$
% \State $\intd{a} \gets \intd{a}_{\text{aux}} + \left( \left( \intd{a}_{l}^{q} \times \bar{\scale}_{\text{out}} \right) \gg (\SFn_{\text{out}} - \SFn_{a}) \right)$
% \item[]
\State $\intd{a} \gets \left( \intd{a}^{q} \times \bar{\scale}_{\text{out}} \right) \gg (\SFn_{\text{out}} - \SFn_{a})$ \Comment{Output rescaling}
\State $\intd{y} \gets \intd{a} + \overline{\bias}$ \Comment{Add bias}
\State $\intw{y} \gets \clipf{\intd{y}, -2^{31}, 2^{31}-1}$ \Comment{Return}
\end{algorithmic}
\end{algorithm}

We introduce the Deployment-Aware Integer Convolution Block, described in Algorithm~\ref{alg:integer_conv}, which is more suitable for hardware deployment than floating-point or fake-quantized convolution, since all operations are implemented using integer arithmetic and bit-shift operations, thereby reducing hardware implementation cost and power dissipation. Notably, the convolution itself is performed at the reduced precision specified by the learned bit widths for activations and weights.

\subsubsection{Notation} 
$v_{\bw}$ denotes any variable represented with bit width $\bw$. Integer variables are indicated by an overbar (e.g., $\bar{v}$), denoting the \texttt{int32} precision, while double overbars (e.g., $\bar{\bar{v}}$) denote double-precision (\texttt{int64}) representations. Moreover, whenever applicable, the superscript $q$ denotes variables that are quantized with learned scales and bit widths.

\subsubsection{Data format for inter-layer communications}
Since quantization parameters (e.g. bit widths, clipping bounds and scale factors) vary across convolutional layers, we adopt a 32-bit integer with $n_a$ bits for the fractional part as the data format for inter-layer data exchange. 

\subsubsection{Offline weight and bias quantization}
Since weights and biases are fixed after QAT or PTQ training, their fixed-point representations can be precomputed and stored. The quantized weights are obtained by~\eqref{eq:quant_intf} as
\begin{equation}
    \label{eq:invc:iw}
    \bar{w}_{\bw_w}^q = \intf{\frac{w}{\scale_w}}{\bw_w},
\end{equation}
where both the bit width $\bw_w$ and scale factor $\scale_w$ are learned by JOMP. In contrast, the bias, usually retained in high precision, is converted into the same data format for inter-layer communications, with $n_a$ fractional bits, so that it can be directly added to obtain the layer output without additional conversion:
\begin{equation}
    \label{eq:invc:ibias}
    \overline{\bias} = \intf{\bias \times 2^{\SFn_a}}{32}.
\end{equation}

\subsubsection{Online activation quantization}
Unlike weight and bias quantization, activation quantization must be performed online with integer arithmetic because the activation value $\bar{x}$ (represented as a 32-bit integer with $n_a$ fractional bits) is content-dependent. 

To quantize further $\bar{x}$ with the learned $\scale_a$ using integer arithmetic, we precompute an integer approximation of $1/\scale_a$:
\begin{equation}
    \label{eq:invc:iscalein}
    \bar{\scale}_{\text{in}} = \intf{\frac{1}{\scale_a} \times 2^{\SFn_{\text{in}}}}{32},
\end{equation}
so that the quantized activation $\bar{x}^{q}_{\bw_a}$ can be obtained in Line 3 of Algorithm~\ref{alg:integer_conv} via integer multiplication, followed by bit-shift and clipping operations.

\subsubsection{Integer convolution}
In our implementation, convolution multiplications are executed at the precision specified by the learned $\bw_a$ and $\bw_w$. The convolution accumulation results are stored in a 64-bit variable $\bar{\bar{a}}^{q}$ (Line 4 in Algorithm~\ref{alg:integer_conv}) to prevent overflow under our current hardware-platform constraints, since the upper bounds of both $\bw_a$ and $\bw_w$ are currently set to 16. Note, however, that customized lower-precision accumulators are feasible when reduced-precision $\bw_a$ and $\bw_w$ are actually employed. 

\subsubsection{Output rescaling and generation}
The output rescaling serves as inverse quantization, compensating for the combined quantization effects of weights and activations in the current layer. This step allows the subsequent layer to adopt a different quantization scheme than the present one. Specifically, the integer convolution result $\intd{a}$ is rescaled using a precomputed 32-bit scale factor $\bar{\scale}_{\text{out}}$ with $\SFn_{\text{out}}$ fractional bits:
\begin{equation}
    \label{eq:invc:iscaleout}
    \bar{\scale}_{\text{out}} = \intf{\scale_a \times \scale_w \times 2^{\SFn_{\text{out}}}}{32}.
\end{equation}
Finally, the bias $\overline{\bias}$ is added and the output is clipped to conform to the data format for inter-layer communications.

Sec.~\ref{sec:integerization} provides further details on how the hyperparameters $n_a$, $n_{in}$, and $n_{out}$, which govern the fractional part of the fixed-point representations at different stages, are determined after JOMP has jointly learned bit widths, clipping bounds and model weights to ensure a balance between numerical precision and overflow safety.

\subsection{Non-convolutional Layers} \label{sec:nonconv_integer}
Beyond convolution, all remaining layers are implemented in integer arithmetic to ensure deterministic inference. All inputs and outputs of these non-convolutional layers adopt the same data format used for inter-layer communications. Moreover, the division in this section is implemented using \textit{integer arithmetic} with rounding to the nearest integer to ensure a close approximation to the floating-point operation.

\subsubsection{LeakyReLU activation function}
In our setting, the slope $\alpha$ of LeakyReLU in the negative domain takes on either 0.1 or 0.01, which rounds to 0 when directly converted into an integer. This degeneration reduces LeakyReLU to ReLU, which is undesirable. To address this, we evaluate $\gamma = \mathrm{round}(1/\alpha)$, which is either 10 or 100, and implement LeakyReLU as follows:
\begin{equation}
    \intw{y} = 
    \begin{cases}
        \intw{x}, & \intw{x} > 0, \\
        \frac{\intw{x}}{\gamma} , & \intw{x} \le 0,
    \end{cases}
\end{equation}
where $\intw{x}$ is the input to the LeakyReLU.

\subsubsection{Non-linear activation functions via lookup tables}
For non-linear functions, such as \textit{sigmoid} $\sigma(\cdot)$, we adopt lookup-table approximations over an input range $[x^{-}, x^{+}]$ with $N$ uniformly sampled entries (e.g., $x^{-}=-16$, $x^{+}=16$, $N=8192$). In practice, inputs are represented in the fixed-point domain scaled by $2^{\SFn_a}$, such that the corresponding integer range becomes
\begin{equation}
    \intw{x}^{-} = x^{-} \times 2^{\SFn_a},
    \quad
    \intw{x}^{+} = x^{+} \times 2^{\SFn_a}.
\end{equation}
Given an integer input $\bar{x}$, the lookup-table index is computed as
\begin{equation}
    \text{idx} =
    \clip \left(
    \frac{(\bar{x} - \intw{x}^{-})(N-1)}
    {\intw{x}^{+} - \intw{x}^{-}},
    \,0,\,N-1 \right)
    .
\end{equation}
The lookup-table entries $\text{LUT}[i]$ are precomputed as 32-bit integers with $n_a$ fractional bits:
\begin{equation}
    \text{LUT}[i]
    =
    \intf{\sigma(x_i)\times 2^{\SFn_a}}{32},
\end{equation}
where
\begin{equation}
    x_i
    =
    x^{-}
    +
    \frac{i(x^{+}-x^{-})}{N-1},
    \quad i=0,\dots,N-1.
\end{equation}

\section{Experimental Setup}

\subsection{Training details}
All codecs are trained on the Vimeo90k~\cite{vimeo} and BVI-DVC~\cite{bvi} datasets following a consistent training protocol~\cite{HyTIP} across all variants to ensure fair comparisons. To construct the performance-complexity trade-off curve, each variant is trained with 4 points, $\lambda_{\text{C}} \in \{0.008, 0.032, 0.128, 0.512\}$. Our best-performing variant, MCR-Hybrid, is trained with 6 points, $\lambda_{\text{C}} \in \{0.008, 0.032, 0.128, 0.512, 1.024, 2.048\}$, for comparison against state-of-the-art methods.
All learnable bit-width parameters are initialized to 16 bits, corresponding to the highest precision in our search space. 
We start from a high-precision configuration that stabilizes early stage of training and prevents premature performance degradation caused by aggressive quantization.
We then perform a joint end-to-end optimization over network weights, clipping bounds $V$ in~\eqref{eq:clippingbounds}, and bit-width parameters $\beta$ in~\eqref{eq:bidwidths}. During training, the bit-width parameters are maintained as continuous variables to enable gradient-based optimization, as mentioned in Sec.~\ref{sec:learn_quant_param_and_bitwidth}. 
After each iteration, a clipping operation is applied to constrain them to the valid range $[8, 16]$.
In practice, the bit-width parameters gradually decrease from the initial 16-bit configuration toward lower-precision regimes due to the BitOPs regularization term.
For the temporal buffer, we adopt a fixed 8-bit quantization to simulate practical deployment scenarios with limited external memory bandwidth and storage.
Specifically, the buffer content is quantized to 8 bits to reduce external memory access, while only the dynamic range parameter is learned and the bit width remains fixed.

\subsection{Testing details}
To evaluate rate-distortion performance, we employ the Bjontegaard Delta-rate (BD-rate)~\cite{bdrate} metric. The evaluation is conducted on four widely used benchmark datasets, including UVG~\cite{uvg}, MCL-JCV~\cite{mcl}, HEVC Class B–E~\cite{hevcctc}, and HEVC-RGB~\cite{hevcrgb}. For each test sequence, the first 96 frames are encoded using an intra period of 32.
Following common practice, we adopt BT.601 for color space conversion between YUV420 and RGB444, with all evaluated codecs operating in the RGB domain.
To avoid padding and ensure fair comparisons, each frame is cropped so that its width and height are multiples of 64. Intra coding is applied at scene cuts across all methods. We report PSNR in the RGB domain and bitrate in bits per pixel (bpp). The BD-rate is computed by averaging per-frame PSNR-RGB and bpp across all frames to generate a dataset-specific rate-distortion point.

\subsection{Complexity metrics} \label{sec:complexity_metrics}
\begin{table}[t]
\centering
\caption{Summary of complexity metrics primarily for decoding.}
\label{tab:compare_metrics}
\begin{tabular}{c|c|c}
\hline
Metric & Description & Platform-Agnostic \\
\hline
Dec. BitOPs & Number of bit operations & \checkmark \\
Dec. WPM & Peak memory usage & \checkmark \\
Dec. Buffer Size & Temporal buffer size & \checkmark \\
Codec Model Size & Codec parameter size & \checkmark \\
Dec. FPS & Frames per second &  $\mathbf{\times}$ \\
\hline
\end{tabular}
\vspace{-1em}
\end{table}

As shown in Table~\ref{tab:compare_metrics}, we primarily adopt platform-agnostic intrinsic complexity metrics, including (i) BitOPs, (ii) weighted peak memory (WPM), (iii) buffer size, (iv) model size, alongside the platform-dependent (v) frames per second (FPS). Our analysis focuses on decoding complexity.

The use of BitOPs is motivated by the fact that although the number of multiply-accumulate operations (MACs) provides a useful estimate of algorithmic complexity, it remains unchanged when the arithmetic precision is varied. For example, two implementations may have the same MAC count, but their hardware characteristics can differ substantially if their modules operate at different bit widths. For implementations targeting ASICs or FPGAs, this distinction is critical, since lower-precision arithmetic units can reduce power dissipation, on-chip area, and even latency by shortening the critical path~\cite{Horowitz2014,fp32_power}. As such, BitOPs is a more appropriate proxy that reflects both the operation count and the arithmetic precision, which together have a direct impact on implementation cost, especially hardware cost. In comparisons that involve both FP models and integer models, we adopt the expedient assumption that FP16 and FP32 models have the same BitOPs as their INT16 and INT32 counterparts, respectively. It should be noted, however, that floating‑point and integer arithmetic are not directly equivalent. At the same bit-width, floating-point operations incur substantially higher hardware cost, power consumption, and computational complexity than their integer counterparts.

WPM quantifies the peak memory usage as the maximum equivalent number of input-resolution feature channels produced by a convolutional layer during decoding. It characterizes the peak memory footprint of the decoder. Note that a high WPM may require larger on-chip memory capacity or more frequent accesses to off-chip memory.

Buffer size denotes the capacity of the temporal buffer, representing the equivalent number of input‑resolution feature channels that must be loaded from external memory when encoding or decoding each frame. A large temporal buffer increases storage cost and memory‑bandwidth demands. Together, WPM and buffer size reveal the memory‑side cost of a neural video codec, which is a critical factor for its practicality.

Codec model size indicates the total number of network parameters of the entire codec (including both the encoder and decoder). Smaller models are generally more suitable for deployment on memory- and resource-constrained devices. When network parameters must be fetched from memory during inference, larger models increase bandwidth demands and energy consumption due to data movement. In contrast, for fully hardwired or on‑chip implementations, this cost is not expressed as runtime loading overhead but rather as requirements on silicon area and storage capacity.

Finally, we report decoding frames-per-second (FPS), following common practice. However, FPS alone is not a sufficiently reliable metric, since it depends not only on the codec design but on numerous implementation- and platform-specific factors. For example, FPS can be influenced by code-level optimization, function-call overhead, memory management, operational complexity, and the interaction with operating system~\cite{compiler_power,dcvcrt}. As a result, two implementations with comparable computational complexity may exhibit substantially different FPS values.

\section{JOMP Experimental Results and Discussions} \label{sec:results}

This section demonstrates the effectiveness of JOMP. 
During training, JOMP learns and assigns precision levels for all the decoding components through fake-quantization, allowing us to examine the impact of integerization across all variants presented in Sec.~\ref{sec:model_variants}. The results regarding the actual integer NVC are presented in Sec.~\ref{sec:integerization}.

\subsection{Bit-width Allocation} 
\label{sec:results_jomp_bw}
\begin{table}
    \caption{Learned bit-width allocation (weights W / activations A) across different coding components under the JOMP framework with $\lambda_{\text{C}} = 0.512$.}
    \centering
    
    \begin{tabular}{c|c|c@{\hspace{5pt}}c|c@{\hspace{5pt}}c|c@{\hspace{5pt}}c}
    \hline
        \multirow{2}{*}{Scheme} &  \multirow{2}{*}{Component} & \multicolumn{2}{c|}{Explicit}   & \multicolumn{2}{c|}{Implicit} & \multicolumn{2}{c}{Hybrid} \\
        & & W & A & W & A & W & A \\ 
        \hline
        \multirow{4}{*}{RC} 
         & Inter-frame main dec. &  9 & 13 & 10 & 13 &  9 & 10 \\
         & Hyperprior dec. & 12 & 14 & 14 & 16 & 12 & 15 \\
         & Prediction network & 10 & 13 & 10 & 14 & 10 & 15 \\
         & Motion dec. & 10 & 14 & 10 & 14 & 10 & 16 \\
         \hline
        \multirow{4}{*}{CC} 
         & Inter-frame main dec. & 10 & 14 & 10 & 14 & 10 & 15 \\
         & Hyperprior dec. & 11 & 12 & 11 & 15 & 11 & 15 \\
         & Prediction network & 10 & 12 & 10 & 13 & 10 & 13 \\
         & Motion dec. & 10 & 12 & 10 & 13 & 10 & 12 \\
          \hline
        \multirow{4}{*}{MCR} 
         & Inter-frame main dec. & 10 & 14 &  9 & 15 &  9 & 15 \\
         & Hyperprior dec. & 11 & 13 & 11 & 16 & 12 & 16 \\
         & Prediction network & 10 & 12 & 10 & 13 & 10 & 14 \\
         & Motion dec. & 10 & 13 & 10 & 14 & 10 & 14 \\
     \hline
    \end{tabular}
    
    \label{tab:results_jomp:precision}
    \vspace{-0.5em}
\end{table}
\begin{figure}[t!]
    \begin{center}
    \includegraphics[width=\linewidth]{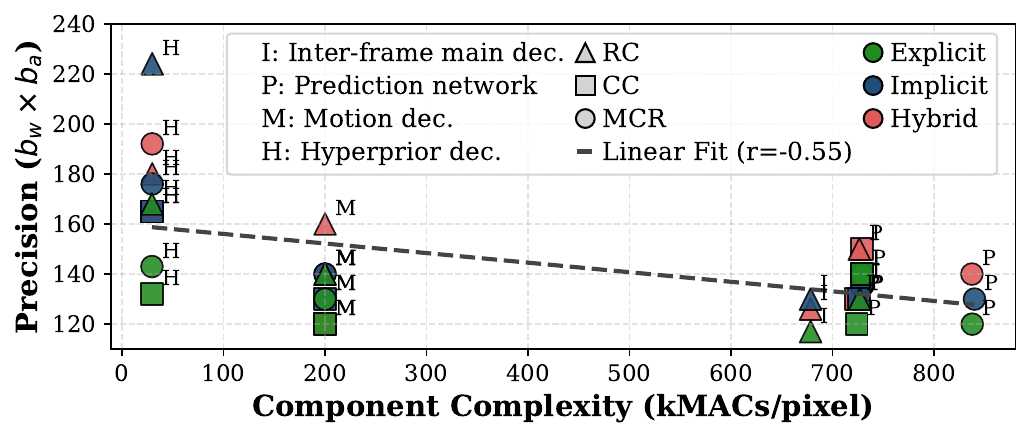}
    \vspace{-2em}
    \caption{Complexity-precision relationship across all model variants at $\lambda_{\text{C}}=0.512$. The x-axis shows the original FP32 component-wise complexity, while the y-axis represents the learned precision measured.
    }
    \label{fig:bit_allocation_complexity}
    \end{center}
    \vspace{-1.5em}
\end{figure}
Firstly, Table~\ref{tab:results_jomp:precision} presents the bit width allocated for both weights (W) and activations (A) of each NVC component spanning across the nine variants and four complexity points.

\textbf{Bit-width allocation varies across different coding components.}
In Table~\ref{tab:results_jomp:precision}, the learned bit widths differ significantly across components, indicating that the model adaptively assigns precision rather than relying on a uniform configuration. 
In Fig.~\ref{fig:bit_allocation_complexity}, this variation shows a clear correlation with component-wise computational complexity. Different components contribute unequally to the overall complexity, which in turn influences the learned bit-width allocation. In the figure, the learned bit widths ($b_w \times b_a$) are plotted against the original FP32 component-wise complexity across all model variants. A clear negative correlation (r = $-0.55$) between component-wise complexity and learned bit width is observed, indicating that components with higher computational cost tend to be assigned lower precision.
In contrast, components with relatively low complexity, such as the hyperprior decoder (H), are typically assigned higher precision, since reducing their bit width yields only marginal gains in overall complexity reduction. These results show that JOMP adapts bit-width allocation according to the complexity of each decoding component.

\subsection{Rate-Distortion and Complexity Results}
\begin{figure*}
    \centering
    \includegraphics[width=0.95\linewidth]{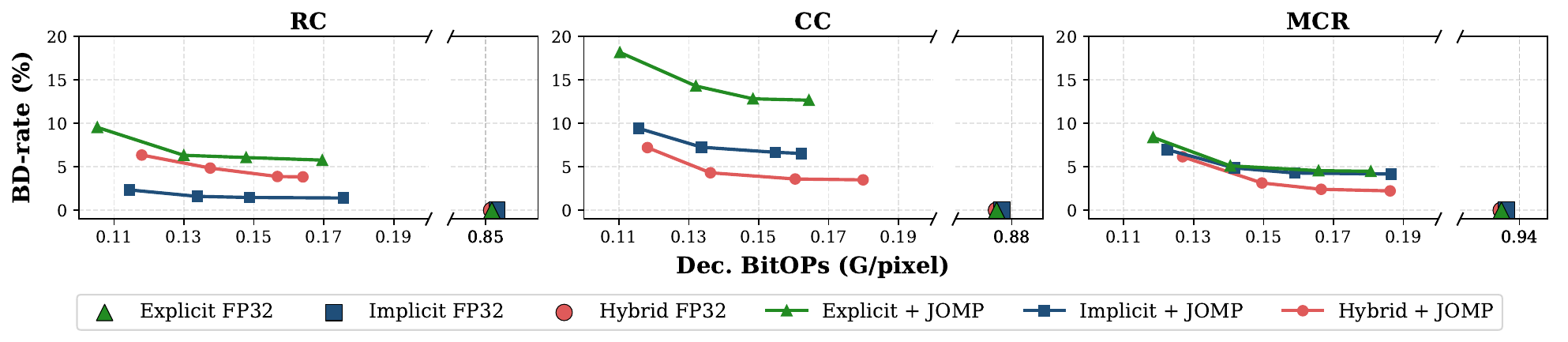}
    \vspace{-1em}
    \caption{BD-rate increase (\%) versus decoder complexity (BitOPs) introduced by JOMP for different model variants. Each point represents the performance gap between the mixed-precision model and its corresponding FP32 baseline, which is used as the anchor for BD-rate evaluation.}
    \label{fig:bd_rate_increase_with_fp}
    \vspace{-1em}
\end{figure*}

Fig.~\ref{fig:bd_rate_increase_with_fp} presents the BD-rate increase due to quantization. The anchor is the respective FP32 NVC variant, which allows us to highlight the JOMP impact over the nine variants. The explicit buffering strategy appears most sensitive to quantization. Nevertheless, aside from ``Explicit + JOMP'' in the CC framework, the BD‑rate increase remains below 10\%, underscoring the efficiency of JOMP.

\begin{figure*}
    \centering
    \includegraphics[width=0.95\linewidth]{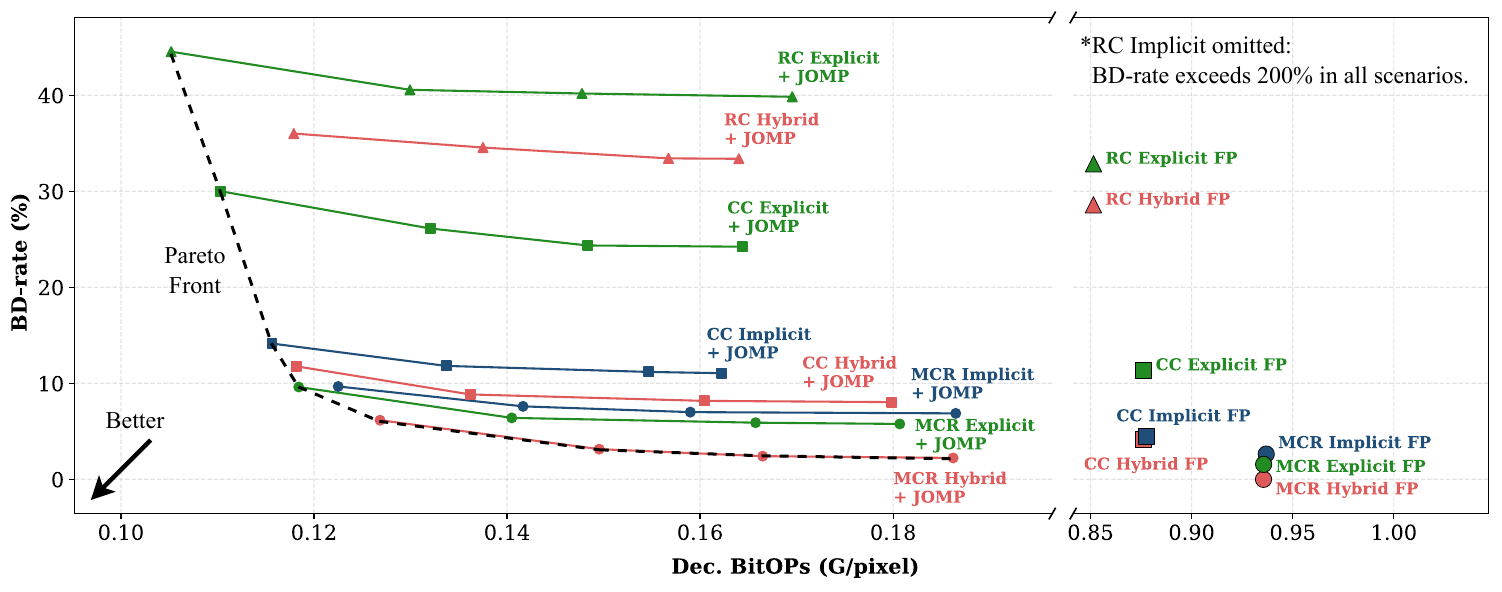}
    \vspace{-1em}
    \caption{BD-rate (\%) versus decoder complexity (BitOPs) for all model variants, including both FP32 and JOMP configurations. The Pareto frontier is highlighted to illustrate the optimal trade-offs. All BD-rate values are computed with respect to the MCR-Hybrid FP32 model as the anchor.}
    \label{fig:bdrate_vs_decops_pareto_jomp}
    \vspace{-0.5em}
    
\end{figure*}

\textbf{Hybrid buffering demonstrates efficiency under quantization.} As shown in Fig.~\ref{fig:bd_rate_increase_with_fp}, this is evidenced by the minimal BD‑rate increase compared with the other two buffering strategies in MCR and CC. Although implicit buffering exhibits a smaller BD‑rate increase in RC, RC variants typically deliver much worse overall rate-distortion performance than MCR and CC variants (see Fig.~\ref{fig:bdrate_vs_decops_pareto_jomp}). 

\textbf{MCR variants demonstrate superior efficiency in balancing coding performance and complexity.} Fig.~\ref{fig:bdrate_vs_decops_pareto_jomp} illustrates the coding performance of different variants under quantization. A clear trend emerges: the integer implementations of MCR variants (MCR Hybrid, MCR Implicit, and MCR Explicit) consistently achieve better rate-distortion performance as compared to the other competing schemes. 
In contrast, the implicit buffer combined with residual coding (RC) leads to severely degraded rate-distortion performance. Generally, the RC variants under-perform relative to the other schemes, particularly in RC-Implicit, where adopting the implicit buffer with residual coding results in pronounced performance degradation. This outcome can be explained by two facts. First, implicit buffering is purely data-driven, and devising an effective training strategy to formulate an effective temporal predictor under this setting remains challenging. Second, with RC (Fig.~\ref{fig:high-level-architecture}(a)), the implicit buffer contains only residual information $F_t$; it is difficult to construct a good temporal predictor based merely on residuals.

\subsection{Comparison with State-of-the-Art NVCs}
\begin{table*}[t]
\centering
\renewcommand{\arraystretch}{1}
\caption{Comparison of complexity and coding performance against state-of-the-art NVC methods. The anchor is VTM 17.0 in Low-delay B mode. FPS is measured on an H100 GPU at a resolution of $1920 \times 1024$.}

\begin{tabular}{l|ccccc|c}
\hline Methods
               & \begin{tabular}[c]{@{}c@{}}Dec. BitOPs\\ (G/pixel)\end{tabular}  &\begin{tabular}[c]{@{}c@{}} Dec. WPM \\ (Channels)\end{tabular}  &  \begin{tabular}[c]{@{}c@{}} Dec. Buffer size\\ (Channels)\end{tabular} & \begin{tabular}[c]{@{}c@{}}Codec Model Size \\ (MB)\end{tabular} & \begin{tabular}[c]{@{}c@{}}Dec. FPS$^{*}$ \\ (on H100)\end{tabular} & \begin{tabular}[c]{@{}c@{}}BD-rate \\ (\%)\end{tabular} \\ \hline

DCVC-FM (FP32)                            & 0.89 & 192 & 51.750 & 69.95 & 6.99 & -19.1 \\
DCVC-RT (FP16)                            & 0.04 & 20  & 2.000  & 39.32 & 195.47 & 5.0 \\ \hline
MCR Hybrid (FP32)                         & 0.95 & 192 & 7.875  & 65.06 & 2.54 & -25.6 \\
MCR Hybrid (QAT, W8A8)                        & 0.06 & 48  & 4.125  & 25.43 & 0.43 & 549.5 \\
MCR Hybrid (QAT, W16A16)                      & 0.24 & 96  & 4.125  & 38.64 & 0.43 & -23.6 \\
MCR Hybrid (JOMP, $\lambda_{\text{C}}=0.512$) & 0.13 & 84  & 4.125  & 30.59 & 0.43 & -23.0 \\
MCR Hybrid (JOMP, $\lambda_{\text{C}}=1.024$) & 0.12 & 78  & 4.125  & 30.27 & 0.43 & -21.5 \\
MCR Hybrid (JOMP, $\lambda_{\text{C}}=2.048$) & 0.11 & 78  & 4.125  & 28.94 & 0.43 & -21.2 \\ \hline
\end{tabular}

\vspace{0.5ex}
\footnotesize
$^{*}$FPS depends on implementation and hardware-specific optimizations. Reported for reference only.
\normalsize

\label{tab:sota_complexity}
\vspace{-1em}

\end{table*}
Table~\ref{tab:sota_complexity} compares the complexity and coding performance of our MCR Hybrid model with recent state-of-the-art learned video codecs. 
We choose MCR Hybrid because it is the strongest among the competing variants. Its implementation is based on a fully integer codec optimized by JOMP rather than fake quantization, ensuring that all reported results reflect practical deployment.

Compared with the FP32 version of MCR Hybrid, JOMP significantly reduces computational complexity while maintaining competitive coding efficiency. 
With $\lambda_{\text{C}}=1.024$, the decoder BitOPs decrease from 0.95 G/pixel to 0.12 G/pixel (a reduction of 87.4\%), while achieving a BD-rate of -21.5\%, which is only 4.1 percentage points higher than the FP32 counterpart (-25.6\%).

Direct comparisons with the integer versions of MobileNVC and DCVC-RT are not included, since these implementations are not publicly available. Therefore, we compare our approach (i) against our own codec under the quantization settings adopted by these methods, and (ii) against the strongest accessible floating-point NVC baselines from related work, namely DCVC-FM and DCVC-RT. This allows us to conduct a fair and reproducible evaluation.

Under the quantization settings adopted in MobileNVC (W8A8), our codec achieves very low complexity (0.06 G/pixel BitOPs), but suffers from severely degraded rate-distortion performance (549.5\% BD-rate). This result indicates that aggressive uniform quantization is insufficient for high-quality neural video coding. In contrast, under the quantization settings adopted in DCVC-RT (W16A16), our model maintains coding performance more effectively (-23.6\% BD-rate) than its W8A8 variant, but with higher complexity (0.24 G/pixel BitOPs). However, comparable rate-distortion performance (-23.0\% BD-rate at $\lambda_{\text{C}}=0.512$) can be achieved by our JOMP-based model with substantially lower complexity (0.13 G/pixel BitOPs). This suggests that JOMP can discover more efficient precision configurations than those selected manually.

When compared with floating-point models, our method also demonstrates favorable performance. Relative to the lightweight DCVC-RT (FP16), it achieves substantially better coding efficiency (-21.2\% vs. 5.0\%) while maintaining moderate computational complexity. Relative to DCVC-FM (FP32), it achieves superior rate-distortion performance while significantly reducing platform-independent complexity metrics.

In terms of decoding throughput, DCVC-RT achieves nearly 200 FPS, whereas the FPS of our current integerized implementation is even lower than that of its FP32 variant. This is because convolution operations are still executed using \texttt{int16} arithmetic on general-purpose GPUs, regardless of the learned bit-widths. Lower-precision computations, such as 10-bit or 12-bit operations, are emulated rather than mapped to native low-precision execution units, which limits the achievable latency gains on current GPU platforms. A similar phenomenon was also reported in~\cite{dcvcrt}.

As noted in Sec.~\ref{sec:complexity_metrics}, FPS is not a sufficient metric in isolation. It depends not only on the codec design but also on a variety of implementation- and platform-specific factors. DCVC-RT was developed specifically for runtime efficiency, emphasizing the reduction of operational overheads. In contrast, our work is primarily motivated by deployment scenarios that may eventually target dedicated hardware, such as ASICs or FPGAs, rather than by maximizing throughput on general-purpose GPUs. Therefore, complexity efficiency is more effectively evaluated through platform-independent metrics, which reflects the computational implications across different compute platforms.

Overall, these results demonstrate that JOMP provides a unified framework for jointly optimizing decoder computation, model size, and memory footprint, enabling favorable performance-complexity trade-offs when integrated into state-of-the-art learned video codecs.

\subsection{Comparison with MP Quantization Frameworks}
\begin{figure}[t!]
    \begin{center}
    \includegraphics[width=\linewidth]{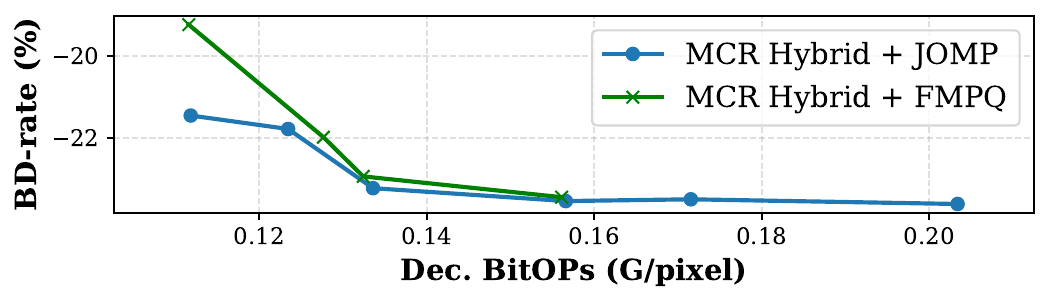}
    \vspace{-2em}
    \caption{Rate-distortion-complexity trade-off curves comparing the JOMP and FMPQ~\cite{immixed} on all datasets. BD-rate is computed relative to the VTM 17.0 (Low-delay B). JOMP achieves superior trade-off solutions, with the performance advantage being most pronounced in the low-complexity regime.}
    \label{fig:mixed_precision}
    \end{center}
    \vspace{-1.5em}
\end{figure}

This section compares JOMP with FMPQ~\cite{immixed}, a recent mixed-precision quantization method originally proposed for neural image codecs. FMPQ is a two-stage, search-based approach: in the first stage, it iteratively applies PTQ to determine the bit width of each module, and in the second stage, it performs QAT to refine both the quantization parameters and the model weights. In contrast, JOMP jointly optimizes bit widths, model weights, and quantization parameters within a single QAT run. For comparison purposes, we apply FMPQ to the MCR-Hybrid model under the same mixed-precision setting. To ensure fairness, both methods are evaluated over the same per-component bit-width search space, namely $[8,16]$.

From Fig.~\ref{fig:mixed_precision}, it is possible to note that our method results in a better Pareto front in the rate-distortion-complexity space compared to FMPQ. Its benefits are particularly evident in the low-BitOPs regime, where our approach yields more favorable trade-offs between coding efficiency and computational cost. This suggests that JOMP is more effective in allocating bit widths under tight compute budgets.

\section{NVC Integerization Results and Discussions}
\label{sec:integerization}
\begin{table*}[t]
\centering
\setlength{\tabcolsep}{3.5pt}

\renewcommand{\arraystretch}{1.2}
% \vspace{-1 em}
\caption{Performance comparison between the fake-quantized model and the integerized model. Results are reported in BD-rate (\%) of MCR Hybrid with $\lambda_{\text{C}}=2.048$, using VTM 17.0 (Low-delay B) as the anchor.}

% \resizebox{\linewidth}{!}{
\begin{tabular}{l|ccccccc|c}
\hline
Dataset                 & UVG    & HEVC-B  & HEVC-C  & HEVC-D    & HEVC-E  & HEVC-RGB  & MCL-JCV   & Average \\ \hline
Fake-Quantized Model        & -29.7  & -19.0   & -13.0   & -33.6     & -12.6   & -20.7     & -21.6     & -21.5   \\
Integerized Model         & -29.3  & -18.9   & -13.0   & -33.5     & -12.2   & -20.5     & -21.2     & -21.2   \\ \hline
\end{tabular}
% }
\label{tab:compare_integer}
\vspace{-1em}

\end{table*}

\begin{table*}[t]
\centering
\caption{Cross-platform consistency validation. Results are \textcolor{black}{obtained from} the UVG and HEVC-B datasets.}
\label{tab:cross_platform}
\begin{tabular}{l|cc|cccccc}
\hline
Model & \multicolumn{2}{c|}{Floating-point Model} & \multicolumn{6}{c}{Integerized JOMP-QAT Model} \\
\hline
 Encoding Platform & H100 & H100 & H100 & H100 & H100 & H100 & H100 & H100 \\

 Decoding Platform & H100 & RTX 4090 & H100 & RTX 4090 & Tesla V100 & Xeon 8480C & Xeon W-2255 & Xeon 6154 \\
\hline

Reconstruction Consistency & $\checkmark$ & $\mathbf{\times}$ & $\checkmark$ & $\checkmark$ & $\checkmark$ & $\checkmark$ & $\checkmark$ & $\checkmark$ \\
\hline
\end{tabular}
\vspace{-1em}

\end{table*}

After training with fake (or simulated) quantization, the model is converted into an integer network for practical deployment. The hyperparameters $n_{in}$, $n_{out}$, and $n_a$ are determined from the observed dynamic ranges of the corresponding floating-point quantities, in order to maximize precision while avoiding overflow. For example, $\SFn_a$ is selected by analyzing the dynamic range of inter-layer data communications within the NVC. In our current implementation, we adopt 32-bit integers (\texttt{int32}) as the format for inter-layer data exchange, where activation values typically fall between $-2^{11}$ and $2^{11}$. Since $\SFn_a$ controls the fractional precision, setting $\SFn_a = 32 - 12 = 20$ allows us to fully utilize the available 32 bits without clipping. The same principle applies to $\SFn_{\text{in}}$ and $\SFn_{\text{out}}$; that is, their values are selected to be as large as possible to preserve numerical precision while remaining within the 32-bit range. In summary, the current settings are $\SFn_a=20$, $\SFn_{\text{in}}=4$, and $\SFn_{\text{out}}=40$.

\subsection{Fake-Quantized Model vs. Integerized Model}
Table~\ref{tab:compare_integer} compares the coding performance of MCR Hybrid between the fake-quantized model and the corresponding integerized model, with JOMP adopting $\lambda_{\text{C}}=2.048$. As shown, the performance gap is negligible across all datasets, with only a 0.3 percentage points difference in the average BD-rate. This result indicates that our fake-quantized model is a reliable proxy for the final integerized implementation. 

\subsection{Cross-platform Consistency Test}

To verify the correctness and determinism of the fully integerized model, we conduct cross-platform experiments, where encoding is performed on NVIDIA H100, while decoding is executed across a diverse set of hardware platforms, including different GPUs and CPUs. 

Table~\ref{tab:cross_platform} summarizes the results for cross-platform reconstruction consistency. Floating-point implementations may introduce reconstruction discrepancies due to platform-dependent numerical behavior. In contrast, our integerized model consistently achieves bit-exact reconstruction across all tested devices. This demonstrates the effectiveness of the proposed integerization method in eliminating cross-platform inconsistencies, making it a suitable solution for practical NVC deployment.

\section{Conclusion}
In this paper, we introduced a Jointly-Optimized Mixed-Precision (JOMP) quantization-aware training framework for neural video codecs. Unlike conventional quantization methods that rely on uniform or heuristic precision settings, JOMP jointly optimizes network weights, quantization scales, and component-wise bit widths under a unified rate-distortion-complexity objective and in an end-to-end manner.

Extensive experiments across multiple coding frameworks (RC, CC, and MCR) and temporal buffering strategies (Explicit, Implicit, and Hybrid) demonstrate that adaptive precision allocation is crucial for balancing coding efficiency and computational cost. We observe that high-complexity components tend to favor lower precision, while lightweight components benefit from higher precision to preserve rate-distortion performance. Overall, JOMP achieves favorable trade-offs between coding performance and decoding complexity by significantly reducing BitOPs while maintaining competitive BD-rate compared to strong floating-point baselines. Notably, when applied to the MCR Hybrid model, JOMP achieves rate-distortion performance comparable with the current state-of-the-art DCVC-FM, while requiring substantially lower decoding complexity. Moreover, the fully integerized implementation enables bit-exact cross-platform inference, addressing a key limitation of existing floating-point NVCs. 

This work bridges the gap between neural video coding and practical deployment by introducing a unified, complexity-aware optimization framework. Future directions include extending the proposed approach to transformer-based codecs, exploring more advanced quantization methods, and investigating hardware-specific optimizations for further efficiency gains in real-world deployment.

\bibliographystyle{IEEEtran}
\bibliography{main}

@String(CVPR  = {IEEE Conf. Comput. Vis. Pattern Recog.})

@String(ICCV  = {Int. Conf. Comput. Vis.})

@String(AAAI  = {AAAI})

@String(ICIP  = {IEEE Int. Conf. Image Process.})

@String(ICASSP=	{ICASSP})

@String(ICME  = {Int. Conf. Multimedia and Expo})

@String(CVPR  = {CVPR})

@String(ICCV  = {ICCV})

@String(ICIP  = {ICIP})

@String(ICME  =	{ICME})

@IEEEtranBSTCTL{IEEEexample:BSTcontrol,
  CTLdash_repeated_names = "no"
}

@String(CVPR= {IEEE Conf. Comput. Vis. Pattern Recog.})

@String(ICCV= {Int. Conf. Comput. Vis.})

@String(ICME = {Int. Conf. Multimedia and Expo})

@String(ICIP = {IEEE Int. Conf. Image Process.})

@String(AAAI = {AAAI})

@INPROCEEDINGS{dvc,
  author={Lu, Guo and Ouyang, Wanli and Xu, Dong and Zhang, Xiaoyun and Cai, Chunlei and Gao, Zhiyong},
  booktitle={2019 IEEE/CVF Conference on Computer Vision and Pattern Recognition (CVPR)}, 
  title={DVC: An End-To-End Deep Video Compression Framework}, 
  year={2019},
  volume={},
  number={},
  pages={10998-11007},
  doi={10.1109/CVPR.2019.01126}}

@INPROCEEDINGS{mlvc,
  author={Lin, Jianping and Liu, Dong and Li, Houqiang and Wu, Feng},
  booktitle={2020 IEEE/CVF Conference on Computer Vision and Pattern Recognition (CVPR)}, 
  title={M-LVC: Multiple Frames Prediction for Learned Video Compression}, 
  year={2020},
  volume={},
  number={},
  pages={3543-3551},
  doi={10.1109/CVPR42600.2020.00360}}

@ARTICLE{nvc,
  author={Liu, Haojie and Lu, Ming and Ma, Zhan and Wang, Fan and Xie, Zhihuang and Cao, Xun and Wang, Yao},
  journal={IEEE Transactions on Circuits and Systems for Video Technology}, 
  title={Neural Video Coding Using Multiscale Motion Compensation and Spatiotemporal Context Model}, 
  year={2021},
  volume={31},
  number={8},
  pages={3182-3196},
  doi={10.1109/TCSVT.2020.3035680}}

@INPROCEEDINGS{ssf,
  author={Agustsson, Eirikur and Minnen, David and Johnston, Nick and Ballé, Jona and Hwang, Sung Jin and Toderici, George},
  booktitle={2020 IEEE/CVF Conference on Computer Vision and Pattern Recognition (CVPR)}, 
  title={Scale-Space Flow for End-to-End Optimized Video Compression}, 
  year={2020},
  volume={},
  number={},
  pages={8500-8509},
  doi={10.1109/CVPR42600.2020.00853}}

@inproceedings{fvc,
  title={FVC: A new framework towards deep video compression in feature space},
  author={Hu, Zhihao and Lu, Guo and Xu, Dong},
  booktitle={Proceedings of the IEEE/CVF Conference on Computer Vision and Pattern Recognition (CVPR)},
  pages={1502--1511},
  year={2021}
}

@inproceedings{dcvc,
  title={Deep contextual video compression},
  author={Li, Jiahao and Li, Bin and Lu, Yan},
  booktitle={Advances in Neural Information Processing Systems},
  volume={34},
  pages={18114--18125},
  year={2021}
}

@inproceedings{canfvc,
  title={CANF-VC: Conditional augmented normalizing flows for video compression},
  author={Ho, Yung-Han and Chang, Chih-Peng and Chen, Peng-Yu and Gnutti, Alessandro and Peng, Wen-Hsiao},
  booktitle={European Conference on Computer Vision},
  pages={207--223},
  year={2022}
}

@article{tcm,
  author={Sheng, Xihua and Li, Jiahao and Li, Bin and Li, Li and Liu, Dong and Lu, Yan},
  journal={IEEE Transactions on Multimedia}, 
  title={Temporal Context Mining for Learned Video Compression}, 
  year={2023},
  volume={25},
  number={},
  pages={7311--7322},
  doi={10.1109/TMM.2022.3220421}}

@inproceedings{hem,
  title={Hybrid spatial-temporal entropy modelling for neural video compression},
  author={Li, Jiahao and Li, Bin and Lu, Yan},
  booktitle={Proceedings of the 30th ACM International Conference on Multimedia (ACM MM)},
  pages={1503--1511},
  year={2022}
}

@inproceedings{dcvcdc,
  title={Neural video compression with diverse contexts},
  author={Li, Jiahao and Li, Bin and Lu, Yan},
  booktitle={Proceedings of the IEEE/CVF Conference on Computer Vision and Pattern Recognition (CVPR)},
  pages={22616--22626},
  year={2023}
}

@INPROCEEDINGS{dcvcfm,
  author={Li, Jiahao and Li, Bin and Lu, Yan},
  booktitle={2024 IEEE/CVF Conference on Computer Vision and Pattern Recognition (CVPR)}, 
  title={Neural Video Compression with Feature Modulation}, 
  year={2024},
  volume={},
  number={},
  pages={26099-26108},
  doi={10.1109/CVPR52733.2024.02466}
}

@ARTICLE{crc,
  author={Brand, Fabian and Seiler, Jürgen and Kaup, André},
  journal={IEEE Transactions on Circuits and Systems for Video Technology}, 
  title={Conditional Residual Coding: A Remedy for Bottleneck Problems in Conditional Inter Frame Coding}, 
  year={2024},
  volume={34},
  number={7},
  pages={6445-6459},
  doi={10.1109/TCSVT.2024.3359948}}

@ARTICLE{maskcrt,
  author={Chen, Yi-Hsin and Xie, Hong-Sheng and Chen, Cheng-Wei and Gao, Zong-Lin and Benjak, Martin and Peng, Wen-Hsiao and Ostermann, Jörn},
  journal={IEEE Transactions on Circuits and Systems for Video Technology}, 
  title={MaskCRT: Masked Conditional Residual Transformer for Learned Video Compression}, 
  year={2024},
  volume={34},
  number={11},
  pages={11980--11992},
  doi={10.1109/TCSVT.2024.3427426}}

@inproceedings{uvg,
author = {Mercat, Alexandre and Viitanen, Marko and Vanne, Jarno},
title = {UVG dataset: 50/120fps 4K sequences for video codec analysis and development},
booktitle = {Proceedings of the 11th ACM Multimedia Systems Conference},
year = {2020},
pages = {297–302},
}

@misc{hevcctc,
  author={Bossen, Frank and others},
  title={Common Test Conditions and Software Reference Configurations},
  note={JCTVC-L1100},
  year={2013}
}

@misc{hevcrgb,
  author={Flynn, David and others},
  title={Common Test Conditions and Software Reference Configurations for HEVC Range Extensions},
  note={JCTVC-N1006},
  year={2013}
}

@inproceedings{mcl,
  title={{MCL-JCV}: a JND-based H. 264/AVC video quality assessment dataset},
  author={Wang, Haiqiang and Gan, Weihao and Hu, Sudeng and Lin, Joe Yuchieh and Jin, Lina and Song, Longguang and Wang, Ping and Katsavounidis, Ioannis and Aaron, Anne and Kuo, C-C Jay},
  booktitle={IEEE International Conference on Image Processing (ICIP)},
  pages={1509--1513},
  year={2016}
}

@article{vimeo,
  author={Xue, Tianfan and Chen, Baian and Wu, Jiajun and Wei, Donglai and Freeman, William T.},
  title={Video Enhancement with Task-Oriented Flow},
  journal={International Journal of Computer Vision},
  volume={127},
  number={8},
  pages={1106--1125},
  year={2019},
  publisher={Springer}
}

@article{bvi,
  author={Ma, Di and Zhang, Fan and Bull, David R.},
  journal={IEEE Transactions on Multimedia}, 
  title={BVI-DVC: A Training Database for Deep Video Compression}, 
  year={2022},
  volume={24},
  number={},
  pages={3847-3858},
  doi={10.1109/TMM.2021.3108943}}

@inproceedings{mobilecodec,
  title={Mobilecodec: neural inter-frame video compression on mobile devices},
  author={Le, Hoang and Zhang, Liang and Said, Amir and Sautiere, Guillaume and Yang, Yang and Shrestha, Pranav and Yin, Fei and Pourreza, Reza and Wiggers, Auke},
  booktitle={Proceedings of the 13th ACM Multimedia Systems Conference},
  pages={324--330},
  year={2022}
}

@INPROCEEDINGS{mobilenvc,
  author={van Rozendaal, Ties and Singhal, Tushar and Le, Hoang and Sautiere, Guillaume and Said, Amir and Buska, Krishna and Raha, Anjuman and Kalatzis, Dimitris and Mehta, Hitarth and Mayer, Frank and Zhang, Liang and Nagel, Markus and Wiggers, Auke},
  booktitle={2024 IEEE/CVF Winter Conference on Applications of Computer Vision (WACV)}, 
  title={MobileNVC: Real-time 1080p Neural Video Compression on a Mobile Device}, 
  year={2024},
  volume={},
  number={},
  pages={4311-4321},
  doi={10.1109/WACV57701.2024.00427}}

@inproceedings{jpegai_deterministic,
  author={Koyuncu, Esin and Solovyev, Timofey and Sauer, Johannes and Alshina, Elena and Kaup, André},
  title={Quantized Decoder in Learned Image Compression for Deterministic Reconstruction},
  booktitle={2024 IEEE International Conference on Acoustics, Speech and Signal Processing (ICASSP)},
  pages={3985--3989},
  year={2024},
  doi={10.1109/ICASSP48485.2024.10448359}
}

@article{aimet,
  author={Nagel, Markus and Fournarakis, Marios and Amjad, Rana Ali and Bondarenko, Yelysei and van Baalen, Mart and Blankevoort, Tijmen},
  title={A White Paper on Neural Network Quantization},
  journal={arXiv preprint arXiv:2106.08295},
  year={2021}
}

@inproceedings{balle_integer,
  author={Ballé, Johannes and Johnston, Nick and Minnen, David},
  title={Integer Networks for Data Compression with Latent-Variable Models},
  booktitle={International Conference on Learning Representations},
  year={2019}
}

@inproceedings{cross_device,
  title={Device interoperability for learned image compression with weights and activations quantization},
  author={Koyuncu, Esin and Solovyev, Timofey and Alshina, Elena and Kaup, Andr{\'e}},
  booktitle={2022 Picture Coding Symposium (PCS)},
  pages={151--155},
  year={2022},
  organization={IEEE}
}

@inproceedings{dcvcrt,
  author={Jia, Zhaoyang and Li, Bin and Li, Jiahao and Xie, Wenxuan and Qi, Linfeng and Li, Houqiang and Lu, Yan},
  title={Towards Practical Real-Time Neural Video Compression},
  booktitle={Proceedings of the IEEE/CVF Conference on Computer Vision and Pattern Recognition (CVPR)},
  pages={12543--12552},
  year={2025},
  doi={10.1109/CVPR52734.2025.01170}
}

@inproceedings{immixed,
  author={Faisal Hossain, Md Adnan and Duan, Zhihao and Zhu, Fengqing},
  title={Flexible Mixed Precision Quantization for Learned Image Compression},
  booktitle={2024 IEEE International Conference on Multimedia and Expo (ICME)},
  pages={1--8},
  year={2024},
  doi={10.1109/ICME57554.2024.10687695}
}

@INPROCEEDINGS{ruhan,
  author={Conceição, Ruhan and Porto, Marcelo and Peng, Wen-Hsiao and Agostini, Luciano},
  booktitle={2025 IEEE International Symposium on Circuits and Systems (ISCAS)}, 
  title={Cross-Platform Neural Video Coding: A Case Study}, 
  year={2025},
  volume={},
  number={},
  pages={1-5},
  doi={10.1109/ISCAS56072.2025.11043556}}

@article{qlic,
  author={Sun, Heming and Yu, Lu and Katto, Jiro},
  journal={IEEE Transactions on Circuits and Systems for Video Technology}, 
  title={Q-LIC: Quantizing Learned Image Compression With Channel Splitting}, 
  year={2025},
  volume={35},
  number={4},
  pages={3798-3811},
  doi={10.1109/TCSVT.2022.3231789}}

@InProceedings{HyTIP,
  author={Chen, Yi-Hsin and Yao, Yi-Chen and Ho, Kuan-Wei and Wu, Chun-Hung and Phung, Huu-Tai and Benjak, Martin and Ostermann, Jörn and Peng, Wen-Hsiao},
  booktitle={2025 IEEE/CVF International Conference on Computer Vision (ICCV)}, 
  title={HyTIP: Hybrid Temporal Information Propagation for Masked Conditional Residual Video Coding}, 
  year={2025},
  volume={},
  number={},
  pages={17889-17898},
  doi={10.1109/ICCV51701.2025.01662}}

@ARTICLE{mp-ptq,
  author={Yu, Jie and Mai, Songping and Zhang, Peng and Jiang, Yucheng and Cheng, Jian},
  journal={IEEE Internet of Things Journal}, 
  title={Mixed-Precision Post-Training Quantization for Learned Image Compression}, 
  year={2025},
  volume={12},
  number={16},
  pages={34392-34405},
  doi={10.1109/JIOT.2025.3578318}}

@misc{bdrate,
  title={Working Practices Using Objective Metrics for Evaluation of Video Coding Efficiency Experiments},
  note={ISO/IEC TR 23002-8, ISO/IEC JTC 1},
  year={2020},
  month={July}
}

@inproceedings{FLIQS,
  author={Dotzel, Jordan and Wu, Gang and Li, Andrew and Umar, Muhammad and Ni, Yun and Abdelfattah, Mohamed S. and Zhang, Zhiru and Cheng, Liqun and Dixon, Martin G. and Jouppi, Norman P. and Le, Quoc V. and Li, Sheng},
  title={FLIQS: One-Shot Mixed-Precision Floating-Point and Integer Quantization Search},
  booktitle={Proceedings of the Third International Conference on Automated Machine Learning},
  series={Proceedings of Machine Learning Research},
  volume={256},
  pages={6/1--6/26},
  year={2024},
  publisher={PMLR}
}

@inproceedings{tenlessons,
  author={Jouppi, Norman P. and Hyun Yoon, Doe and Ashcraft, Matthew and Gottscho, Mark and Jablin, Thomas B. and Kurian, George and Laudon, James and Li, Sheng and Ma, Peter and Ma, Xiaoyu and Norrie, Thomas and Patil, Nishant and Prasad, Sushma and Young, Cliff and Zhou, Zongwei and Patterson, David},
  booktitle={2021 ACM/IEEE 48th Annual International Symposium on Computer Architecture (ISCA)}, 
  title={Ten Lessons From Three Generations Shaped Google’s TPUv4i : Industrial Product}, 
  year={2021},
  volume={},
  number={},
  pages={1-14},
  doi={10.1109/ISCA52012.2021.00010}}

@inproceedings{fracbits,
  title={FracBits: Mixed Precision Quantization via Fractional Bit-Widths},
  author={Linjie Yang and Qing Jin},
  booktitle={AAAI Conference on Artificial Intelligence},
  year={2020}
}

@inproceedings{SDQ,
  author={Huang, Xijie and Shen, Zhiqiang and Li, Shichao and Liu, Zechun and Hu, Xianghong and Wicaksana, Jeffry and Xing, Eric P. and Cheng, Kwang-Ting},
  title={SDQ: Stochastic Differentiable Quantization with Mixed Precision},
  booktitle={Proceedings of the 39th International Conference on Machine Learning},
  volume={162},
  pages={9295--9309},
  year={2022}
}

@InProceedings{DDQ,
  title = 	 {Differentiable Dynamic Quantization with Mixed Precision and Adaptive Resolution},
  author =       {Zhang, Zhaoyang and Shao, Wenqi and Gu, Jinwei and Wang, Xiaogang and Luo, Ping},
  booktitle = 	 {Proceedings of the 38th International Conference on Machine Learning},
  pages = 	 {12546--12556},
  year = 	 {2021},
  editor = 	 {Meila, Marina and Zhang, Tong},
  volume = 	 {139},
  series = 	 {Proceedings of Machine Learning Research},
  month = 	 {18--24 Jul},
  publisher =    {PMLR}
}

@inproceedings{uhlich2020,
  author={Uhlich, Stefan and Mauch, Lukas and Cardinaux, Fabien and Yoshiyama, Kazuki and Garcia, Javier Alonso and Tiedemann, Stephen and Kemp, Thomas and Nakamura, Akira},
  title={Mixed Precision DNNs: All You Need Is a Good Parametrization},
  booktitle={International Conference on Learning Representations},
  year={2020},
}

@article{ste,
  title={Estimating or propagating gradients through stochastic neurons for conditional computation},
  author={Bengio, Yoshua and L{\'e}onard, Nicholas and Courville, Aaron},
  journal={arXiv preprint arXiv:1308.3432},
  year={2013}
}

@INPROCEEDINGS{Horowitz2014,
  author={Horowitz, Mark},
  booktitle={2014 IEEE International Solid-State Circuits Conference Digest of Technical Papers (ISSCC)}, 
  title={1.1 Computing's energy problem (and what we can do about it)}, 
  year={2014},
  volume={},
  number={},
  pages={10-14},
  doi={10.1109/ISSCC.2014.6757323}}

@ARTICLE{fp32_power,
  author={Mao, Wei and Li, Kai and Cheng, Quan and Dai, Liuyao and Li, Boyu and Xie, Xinang and Li, He and Lin, Longyang and Yu, Hao},
  journal={IEEE Transactions on Very Large Scale Integration Systems}, 
  title={A Configurable Floating-Point Multiple-Precision Processing Element for HPC and AI Converged Computing}, 
  year={2022},
  volume={30},
  number={2},
  pages={213-226},
  doi={10.1109/TVLSI.2021.3128435}}

@inproceedings{compiler_power,
author = {Georgiou, Kyriakos and Blackmore, Craig and Xavier-de-Souza, Samuel and Eder, Kerstin},
title = {Less is More: Exploiting the Standard Compiler Optimization Levels for Better Performance and Energy Consumption},
year = {2018},
isbn = {9781450357807},
address = {New York, NY, USA},
doi = {10.1145/3207719.3207727},
booktitle = {Proceedings of the 21st International Workshop on Software and Compilers for Embedded Systems},
pages = {35–42},
numpages = {8},
location = {Sankt Goar, Germany},
}

@ARTICLE{jpegai,
  author={Esenlik, Semih and Wu, Yaojun and Zhang, Zhaobin and Wang, Ye-Kui and Zhang, Kai and Zhang, Li and Ascenso, João and Liu, Shan},
  journal={IEEE Transactions on Circuits and Systems for Video Technology}, 
  title={An Overview of the JPEG AI Learning-Based Image Coding Standard}, 
  year={2026},
  volume={36},
  number={2},
  pages={2520-2537},
  doi={10.1109/TCSVT.2025.3613244}}

@ARTICLE{fpxnic,
  author={Jia, Chuanmin and Hang, Xinyu and Wang, Shanshe and Wu, Yaqiang and Ma, Siwei and Gao, Wen},
  journal={IEEE Transactions on Circuits and Systems for Video Technology}, 
  title={FPX-NIC: An FPGA-Accelerated 4K Ultra-High-Definition Neural Video Coding System}, 
  year={2022},
  volume={32},
  number={9},
  pages={6385-6399},
  doi={10.1109/TCSVT.2022.3164059}}

@INPROCEEDINGS{c3,
  author={Kim, Hyunjik and Bauer, Matthias and Theis, Lucas and Schwarz, Jonathan Richard and Dupont, Emilien},
  booktitle={2024 IEEE/CVF Conference on Computer Vision and Pattern Recognition (CVPR)}, 
  title={C3: High-Performance and Low-Complexity Neural Compression from a Single Image or Video}, 
  year={2024},
  volume={},
  number={},
  pages={9347-9358},
  doi={10.1109/CVPR52733.2024.00893}}

@inproceedings{hinerv,
 author = {Kwan, Ho Man and Gao, Ge and Zhang, Fan and Gower, Andrew and Bull, David},
 booktitle = {Advances in Neural Information Processing Systems},
 pages = {72692--72704},
 title = {HiNeRV: Video Compression with Hierarchical Encoding-based Neural Representation},
 volume = {36},
 year = {2023}
}

\vspace{-4em}
\begin{IEEEbiography}[{\includegraphics[width=1in,height=1.25in,clip,keepaspectratio]{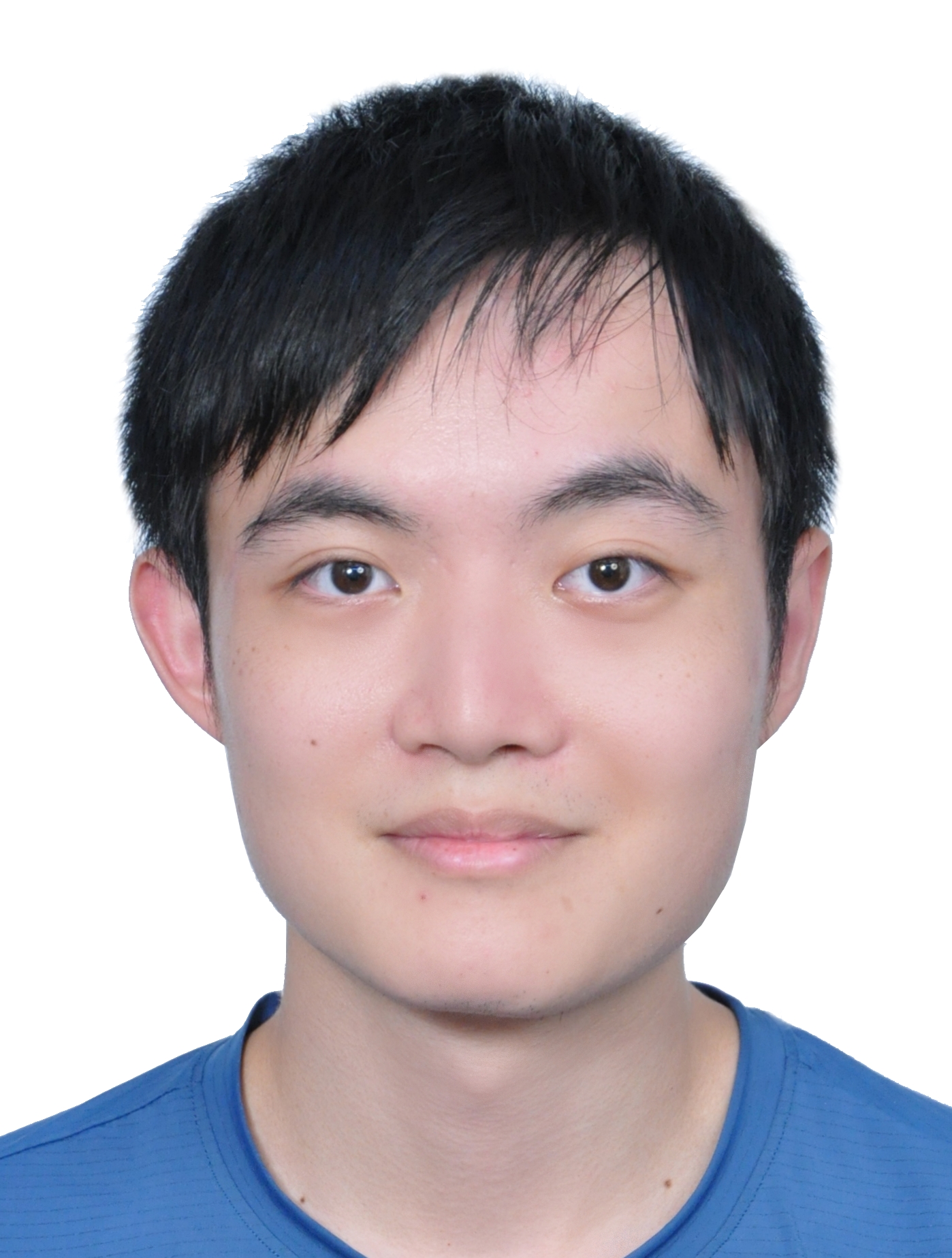}}]{Yu-Hsiang Lin}
received his B.S. degree in applied Computer Science and Information Engineering from National Chung Cheng University (CCU), Taiwan, in 2024. He is currently pursuing his M.S. degree in Computer Science, National Yang Ming Chiao Tung University (NYCU), Taiwan. His research interests include learning-based image/video coding, computer vision, and deep learning.
\end{IEEEbiography}
\vspace{-4em}

\begin{IEEEbiography}[{\includegraphics[width=1in,height=1.25in,clip,keepaspectratio]{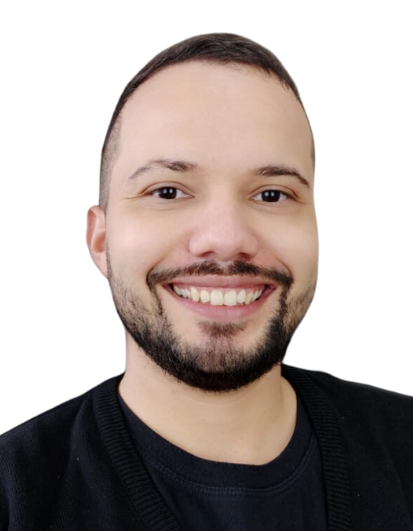}}]{Ruhan Conceição} holds a Master’s degree in Computer Science (2015) and a Bachelor's degree in Computer Engineering (2014) from the Federal University of Pelotas, Brazil, and is currently pursuing a Ph.D. in Computer Science at the same institution. In 2024, he was a visiting Ph.D. researcher at National Yang Ming Chiao Tung University, Taiwan. He is a Professor at the Federal Institute of Education, Science, and Technology Sul-rio-grandense, where he coordinated the Integrated Technical Program in Internet Informatics from 2019 to 2022. His research interests include hardware design and complexity-aware software solutions for traditional video codecs, light-field imaging, and neural network-based video codecs.
\end{IEEEbiography}
\vspace{-4em}

\begin{IEEEbiography}[{\includegraphics[width=1in,height=1.25in,clip,keepaspectratio]{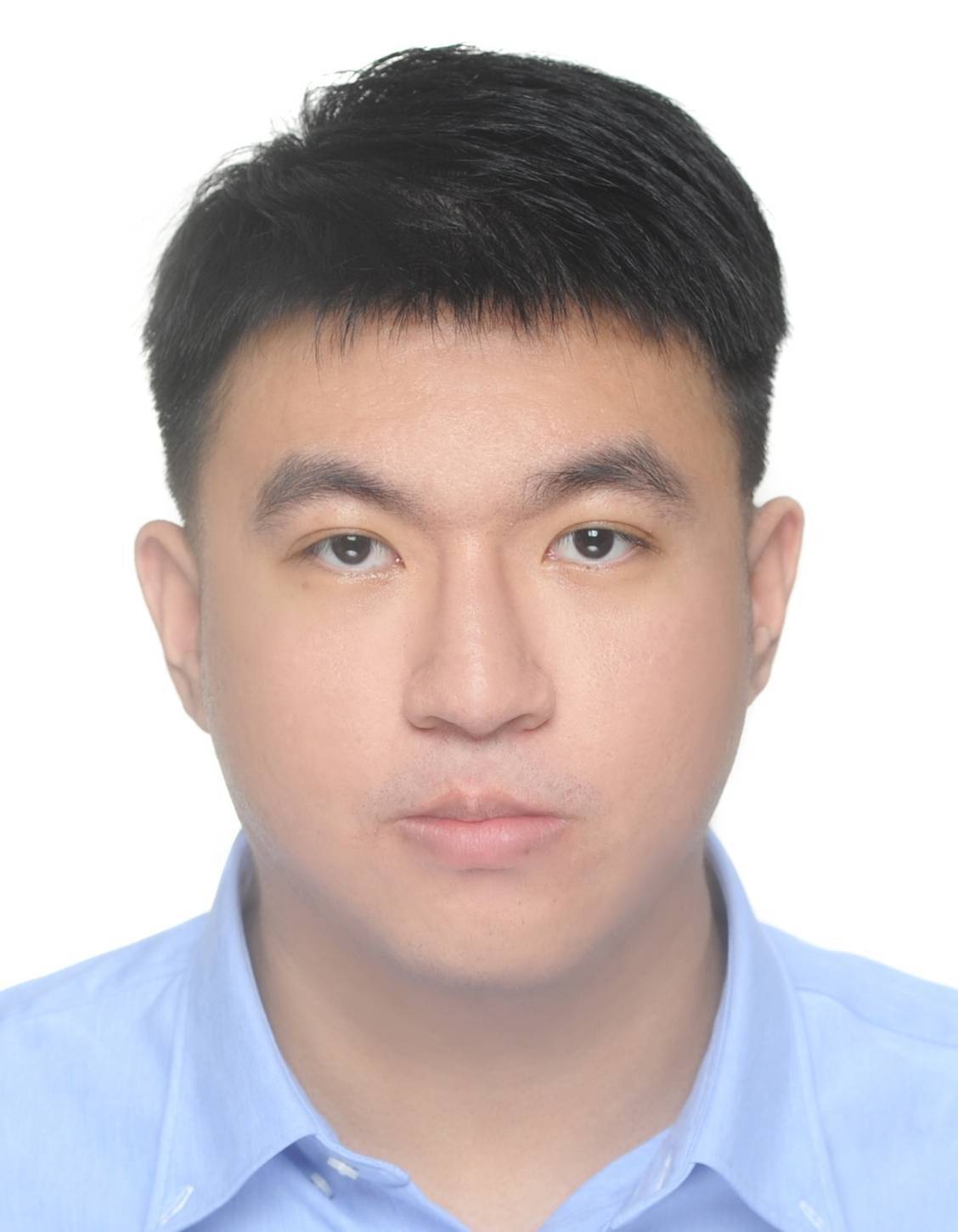}}]{Chun-Hung Wu} received his B.S. degree in applied compute science from National Yang Ming Chiao Tung University (NYCU), Taiwan, in 2024. He is currently pursuing his M.S. degree in Computer Science at National Yang Ming Chiao Tung University (NYCU), Taiwan. His research interests include image/video compression, computer vision.

\end{IEEEbiography}
\vspace{-4em}

\begin{IEEEbiography}[{\includegraphics[width=1in,height=1.25in,clip,keepaspectratio]{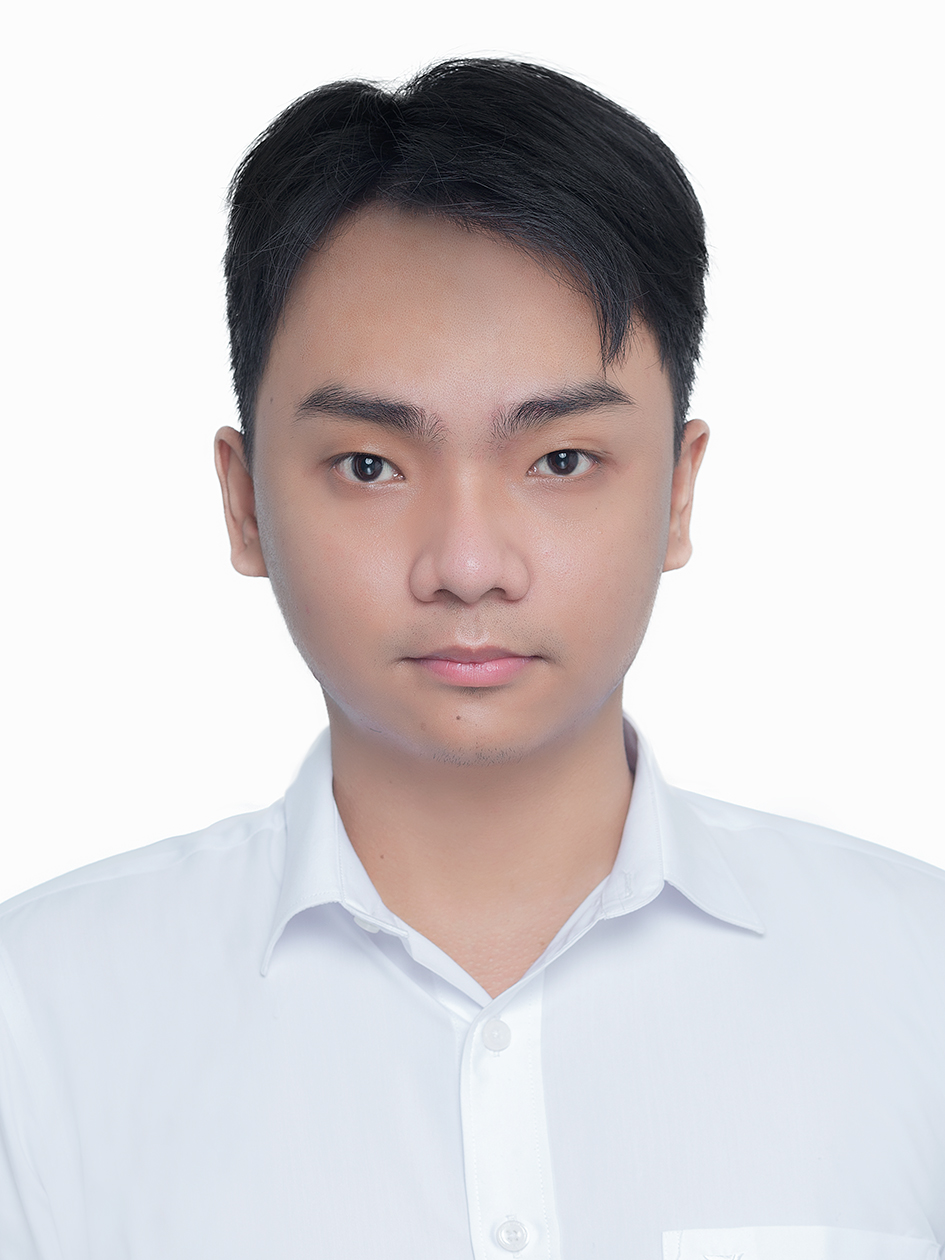}}]{Huu-Tai Phung} is currently pursuing a Ph.D. in Computer Science at National Yang Ming Chiao Tung University, Taiwan. He received his Master’s degree in Electrical Engineering from National Chung Cheng University, Taiwan, and his Bachelor’s degree in Software Engineering from the University of Danang, Vietnam, in 2021. His research interests include neural image and video compression, vision foundation models, and visual quality enhancement.

\end{IEEEbiography}
\vspace{-4em}

\begin{IEEEbiography}[{\includegraphics[width=1in,height=1.25in,clip,keepaspectratio]{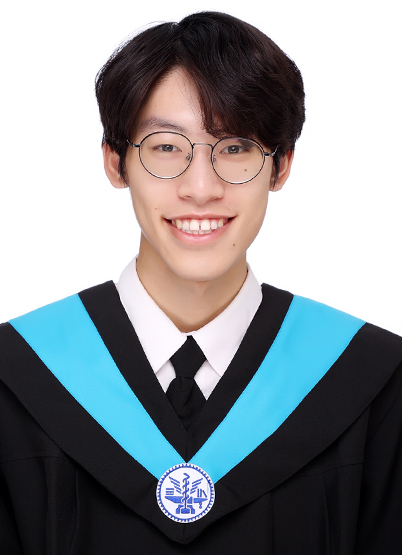}}]{Tzu-Hsiang Chou}
received his B.S. degree in Computer Science from National Yang Ming Chiao Tung University (NYCU), Taiwan, in 2025. He is currently pursuing his M.S. degree in Computer Science at NYCU, Taiwan. His research interests include image/video compression and physics-informed artificial intelligence (physics AI).
\end{IEEEbiography}
\vspace{-4em}

\begin{IEEEbiography}[{\includegraphics[width=1in,height=1.25in,clip,keepaspectratio]{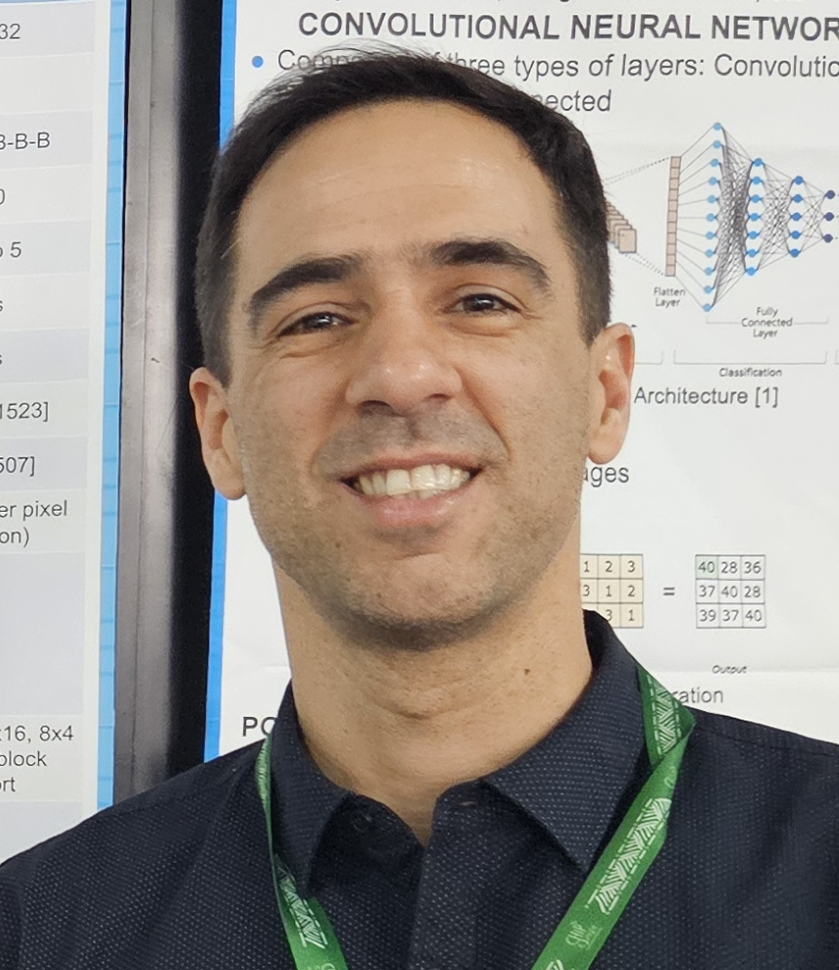}}]{Marcelo Porto} (Senior Member, IEEE) received the M.S. and Ph.D. degrees in computer science from Federal University of Rio Grande do Sul (UFRGS), Brazil, in 2008 and 2012, respectively. He is currently a Professor with the Federal University of Pelotas (UFPel), Brazil, and a member of the Video Technology Research Group (ViTech). He is a permanent member of the Graduate Program in Computing (PPGC) of UFPel. Dr. Porto is a Senior Member of IEEE, and a Member of SBC and SBMicro Brazilian societies. He holds the status of CNPq (National Council for Scientific and Technological Development) Brazilian Distinguished Researcher through a PQ-1D grant. His research interests include video coding, motion estimation algorithms, point cloud compression, coding complexity reduction, and energy-efficient VLSI design for video coding.

\end{IEEEbiography}
\vspace{-4em}

\begin{IEEEbiography}[{\includegraphics[width=1in,height=1.25in,clip,keepaspectratio]{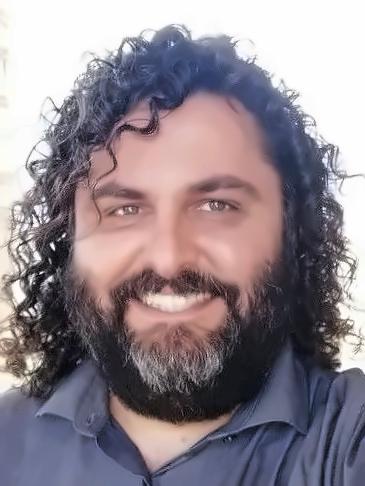}}]{Luciano Volcan Agostini} (Senior Member, IEEE) received his M.S. and Ph.D. degrees in Computer Science from the Federal University of Rio Grande do Sul, Brazil, in 2002 and 2007, respectively. Since 2002, he has been a professor at the Federal University of Pelotas (UFPel), Brazil, where he is currently a Full Professor. Since 2013, he has held the status of CNPq Productivity Research Fellow. He is also the Vice President of the Brazilian Microelectronics Society (SBMicro) and leads the Video Technology Research Group at UFPel. From 2013 to 2017, he served as Vice President for Research and Graduate Studies at UFPel. Additionally, he is a member of the Advisory Committees of CNPq and FAPERGS, two of Brazil’s leading research funding agencies. Dr. Agostini has authored over 300 papers published in prestigious international journals and conferences. He is an Associate Editor for IEEE TCSVT and IEEE OJCS. He is also a Senior Member of ACM and a member of the SBC and SBMicro societies in Brazil.
\end{IEEEbiography}
\vspace{-4em}

\begin{IEEEbiography}[{\includegraphics[width=1in,height=1.25in,clip,keepaspectratio]{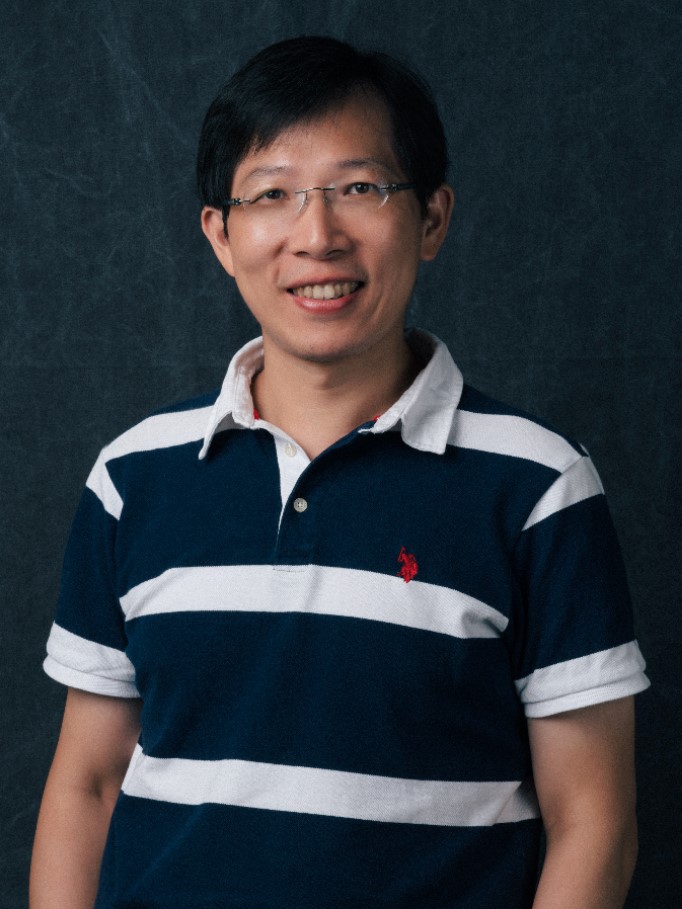}}]{Wen-Hsiao Peng} (M’09-SM’13-F’25) received his Ph.D. degree from National Chiao Tung University (NCTU), Taiwan, in 2005. He was with the Intel Microprocessor Research Laboratory, USA, from 2000 to 2001. Since 2003, he has actively participated in the ISO/IEC and ITU-T video coding standardization process and contributed to the development of H.264/AVC Scalable Amendment, H.265/HEVC, H.265/HEVC Screen Content Coding Extensions (SCC), H.266/VVC, and JPEG AI. He was a Visiting Scholar with the IBM Thomas J. Watson Research Center, USA, from 2015 to 2016. He has authored over 120 journal/conference papers and over 60 ISO/IEC and ITU-T standards contributions. Dr. Peng was Chair of the IEEE Circuits and Systems Society (CASS) Visual Signal Processing (VSPC) Technical Committee from 2020 to 2022. He was Distinguished Lecturer of IEEE CASS (2022-2023) and of APSIPA (2017-2018). He was appointed Editor-in-Chief of the IEEE Journal on Emerging and Selected Topics in Circuits and Systems (JETCAS) for 2024-2025. Dr. Peng is a Fellow of the Higher Education Academy (FHEA), and a Fellow of the IEEE. 
\end{IEEEbiography}

% Supplementary Material
\newpage
\twocolumn[
\begin{center}
{\LARGE JOMP: Jointly-Optimized Mixed-Precision Quantization Across Neural Video Coding Frameworks and Buffering Strategies\par}
\vspace{1em}
{Supplementary Material \par}
\vspace{1em}
\end{center}
]

\section{Derivation of Gradients}
This section provides the detailed derivation of the gradients used in the optimization framework.

During forward propagation, the corresponding equations are given by
\begin{equation*}
    b = \round{\bwreal}
\end{equation*}
\begin{equation*}
    \qbmin_b, \qbmax_b = -2^{b-1}, 2^{b-1}-1
\end{equation*}
\begin{equation*}
    s = \frac{\vmax}{\qbmax_b}
\end{equation*}
\begin{equation*}
    \hat{v} = \round{\clip(\frac{v}{s},\,\qbmin_b,\,\qbmax_b)} \times s = \round{\clip(u,\,\qbmin_b,\,\qbmax_b)} \times s
\end{equation*}

\subsection{Gradient w.r.t. Continuous Clipping Bound $\vmax$}
\subsubsection{Case 1: $\qbmin_b < u < \qbmax_b$}
\begin{equation*}
    \hat{v} = \round{u} \cdot s
\end{equation*}
\begin{align*}
    \frac{\partial \hat{v}}{\partial \vmax} & = \round{u} \cdot \frac{\partial s}{\partial \vmax} + s \cdot \frac{\partial \round{u}}{\partial \vmax} \\
    & = \round{u} \cdot \frac{\partial}{\partial \vmax} \left(\frac{\vmax}{\qbmax_b}\right) + s \cdot \frac{\partial}{\partial \vmax} \left(v \cdot \frac{\qbmax_b}{\vmax}\right) \\
    & = \round{u} \cdot \frac{1}{\qbmax_b} - s \cdot \frac{v \cdot \qbmax_b}{(\vmax)^2} \\
    & = \frac{\round{u}}{\qbmax_b} - \frac{u}{\qbmax_b} \\
    & = \frac{\round{u} - u}{\qbmax_b}
\end{align*}

\subsubsection{Case 2: $u \le \qbmin_b$}
\begin{align*}
\hat{v}
=
\qbmin_b \cdot s
=
\qbmin_b \cdot \frac{\vmax}{\qbmax_b}
\end{align*}

\begin{align*}
\frac{\partial \hat{v}}{\partial \vmax} &= \frac{\qbmin_b}{\qbmax_b}
\end{align*}

\subsubsection{Case 3: $u \ge \qbmax_b$}
\begin{align*}
\hat{v}
=
\qbmax_b \cdot s
=
\qbmax_b \cdot \frac{\vmax}{\qbmax_b} = \vmax
\end{align*}

\begin{align*}
\frac{\partial \hat{v}}{\partial \vmax} &= 1
\end{align*}

\subsection{Gradient w.r.t. Continuous Bit Width $\bwreal$}
\subsubsection{Case 1: $\qbmin_b < u < \qbmax_b$}
\begin{equation*}
    \hat{v} = \round{u} \cdot s
\end{equation*}
\begin{align*}
    \frac{\partial \hat{v}}{\partial \bwreal} & = \round{u} \cdot \frac{\partial s}{\partial \bwreal} + s \cdot \frac{\partial \round{u}}{\partial \bwreal} \\
    & = (v-\hat{v}) \times \frac{2^{\bwreal - 1}\ln 2}{\qbmax_{\bw}}
\end{align*}

where
\begin{align*}
    \round{u} \cdot \frac{\partial s}{\partial \bwreal} &= \round{u} \cdot \frac{\partial}{\partial \bwreal} \left(\frac{\vmax}{\qbmax_b}\right) \\
    & = \round{u} \cdot \vmax \cdot (-1)(\qbmax_b)^{-2} \cdot \frac{\partial \qbmax_b}{\partial \bwreal} \\
    & = \round{u} \cdot \frac{\vmax}{\qbmax_b} \cdot \frac{-1}{\qbmax_b} \cdot \frac{\partial \qbmax_b}{\partial \bwreal} \\
    & = \frac{-\hat{v}}{\qbmax_b} \cdot \frac{\partial \qbmax_b}{\partial \bwreal},
\end{align*}

\begin{align*}
    s \cdot \frac{\partial \round{u}}{\partial \bwreal} & = \frac{\vmax}{\qbmax_b} \cdot \frac{\partial}{\partial \bwreal} \left(v \cdot \frac{\qbmax_b}{\vmax}\right) \\
    & = \frac{v}{\qbmax_b} \cdot \frac{\partial \qbmax_b}{\partial \bwreal},
\end{align*}

and
\begin{align*}
    \frac{\partial \qbmax_b}{\partial \bwreal} &= 2^{\bwreal - 1}\ln 2.
\end{align*}

\subsubsection{Case 2: $u \le \qbmin_b$}
\begin{align*}
\hat{v}
=
\qbmin_b \cdot s
=
\qbmin_b \cdot \frac{\vmax}{\qbmax_b}
\end{align*}

\begin{align*}
\frac{\partial \hat{v}}{\partial \bwreal} &= \frac{\partial}{\partial \bwreal} \left(\vmax \frac{\qbmin_b}{\qbmax_b}\right) \\
& = \vmax \cdot \frac{\qbmax_b \cdot \frac{\partial \qbmin_b}{\partial \bwreal} - \qbmin_b \cdot \frac{\partial \qbmax_b}{\partial \bwreal}}{(\qbmax_b)^2} \\
& = \vmax \cdot \frac{\qbmax_b \cdot (-2^{\bwreal - 1}\ln 2) - \qbmin_b \cdot (2^{\bwreal - 1}\ln 2)}{(\qbmax_b)^2} \\
& = \vmax \cdot \frac{(-\qbmax_b - \qbmin_b) \cdot 2^{\bwreal - 1}\ln 2}{(\qbmax_b)^2} \\
& = \vmax \cdot \frac{2^{\bwreal - 1}\ln 2}{(\qbmax_b)^2}
\end{align*}

\subsubsection{Case 3: $u \ge \qbmax_b$}
\begin{align*}
\hat{v}
=
\qbmax_b \cdot s
=
\qbmax_b \cdot \frac{\vmax}{\qbmax_b} = \vmax
\end{align*}

\begin{align*}
\frac{\partial \hat{v}}{\partial \bwreal} &= 0
\end{align*}

\section{Additional Rate-Distortion Curves}
\begin{figure*}[t!]
    \begin{center}
    \includegraphics[width=\linewidth]{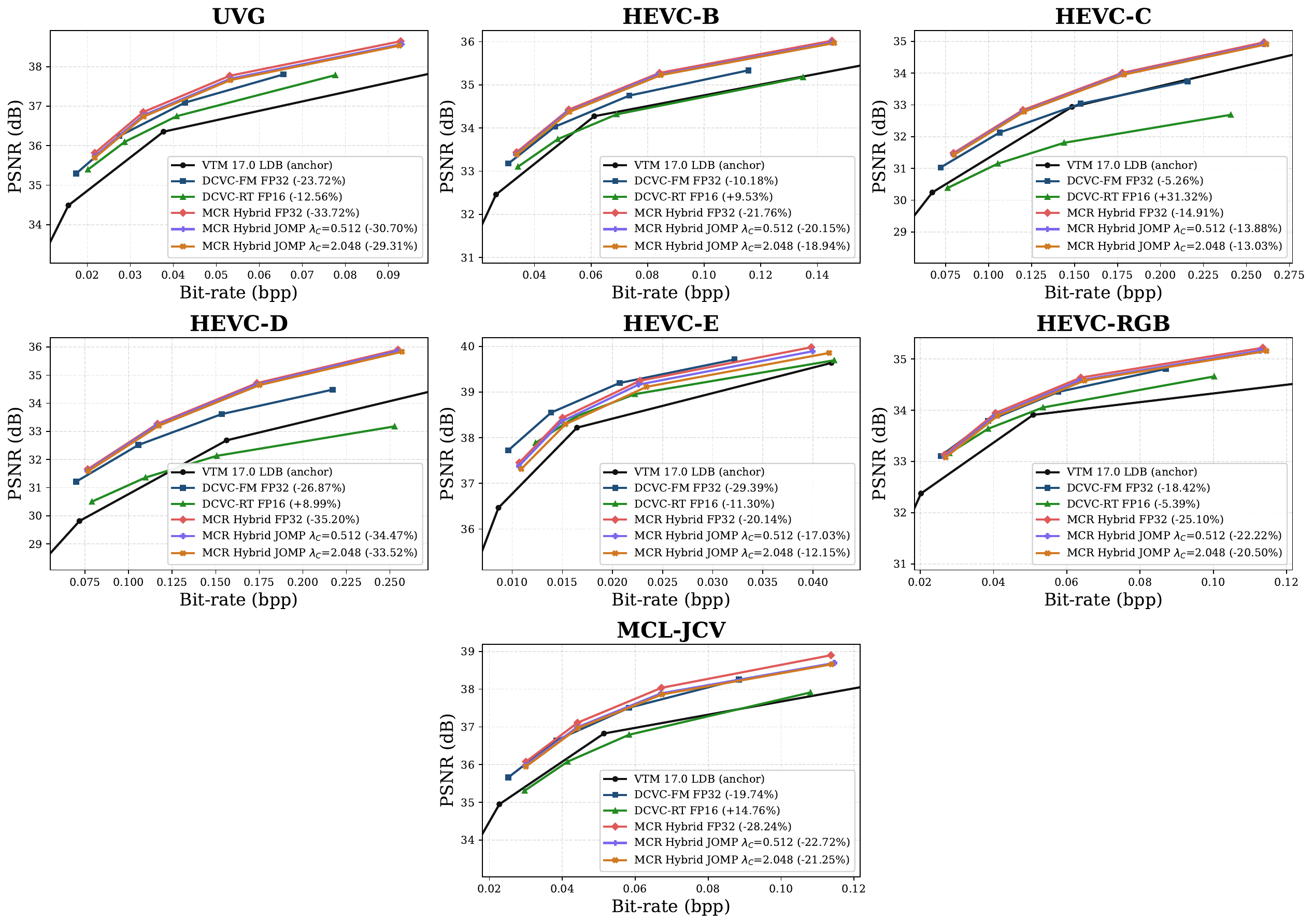}
    \caption{Rate-distortion comparison with the state-of-the-art NVCs. The anchor is VTM 17.0 (Low-delay B).}
    \label{fig:rd_curve}
    \end{center}
    % \vspace{-3 em}
\end{figure*}

For completeness, Fig.~\ref{fig:rd_curve} presents the full rate-distortion curves of the evaluated methods.

\section{Visualization for Cross-platform Consistency}
\begin{figure}[t!]
    \begin{center}
    \includegraphics[width=0.9\linewidth]{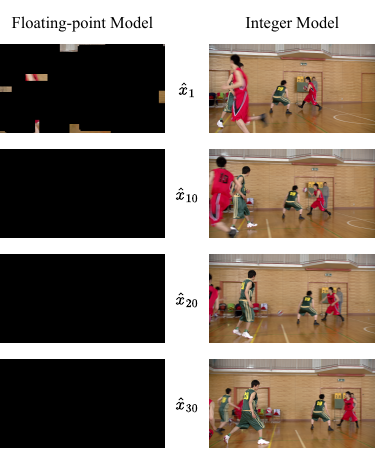}
    \caption{Visual comparison of reconstructed frames from the {\it BasketballDrive} sequence under floating-point and integer implementations. The sequence is encoded on an NVIDIA H100 and decoded on an NVIDIA RTX 4090.}
    \label{fig:cross_platform}
    \end{center}

    \vspace{-1em}

\end{figure}
\begin{figure}[t!]
    \begin{center}
    \includegraphics[width=0.9\linewidth]{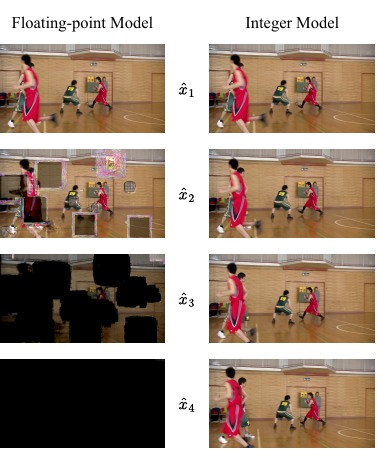}
    \caption{Visual comparison of reconstructed frames from the {\it BasketballDrive} sequence under floating-point and integer implementations. Unlike Fig.~\ref{fig:cross_platform}, the first frame is correctly reconstructed in this example, allowing the temporal accumulation of cross-platform numerical discrepancies to be more clearly observed. The sequence is encoded on an NVIDIA H100 and decoded on an NVIDIA RTX 4090.}
    \label{fig:cross_platform_IntraCorrect}
    \end{center}

    \vspace{-1em}

\end{figure}
Fig.~\ref{fig:cross_platform} further provides a visual comparison using selected frames from the {\it BasketballDrive} sequence, showing the reconstructed outputs from both floating-point and integer models. The floating-point model fails to reconstruct the input sequence properly.

Fig.~\ref{fig:cross_platform_IntraCorrect} shows a case where the first frame is reconstructed correctly. Nevertheless, small cross-platform numerical discrepancies accumulate through temporal prediction and buffered references, leading to progressively increasing reconstruction errors in subsequent frames.

\end{document}